\documentclass[11pt]{article}
\pdfoutput=1
\usepackage{jheppub}
\usepackage[utf8]{inputenc}
\usepackage[english]{babel}
\usepackage{amsmath,amssymb,amsbsy,amstext,amsthm,booktabs,slashed,graphicx,amsfonts,xcolor,subcaption}
\usepackage{multirow,bm,enumerate,longtable,array,colortbl}
\usepackage{hyperref}
\usepackage[capitalise,noabbrev,nameinlink]{cleveref}
\usepackage{mathtools}
\usepackage{listings}
\usepackage{float}
\usepackage{tikz}
\usepackage[normalem]{ulem}
\usepackage[section]{placeins}   
\usetikzlibrary{arrows.meta,positioning,shapes.geometric,calc,decorations.pathmorphing,decorations.markings}
\tikzset{
  ferm/.style={draw, thick, postaction={decorate},
    decoration={markings, mark=at position 0.58 with {\arrow{Latex[length=1.6mm]}}}},
  fermb/.style={draw, thick, postaction={decorate},
    decoration={markings, mark=at position 0.58 with {\arrowreversed{Latex[length=1.6mm]}}}},
  vec/.style={draw, thick, decorate,
    decoration={snake, amplitude=1.1pt, segment length=4.6pt, pre length=1pt, post length=1pt}},
  glu/.style={draw, thick, decorate,
    decoration={coil, amplitude=2.1pt, segment length=3.4pt, aspect=0.75, pre length=1pt, post length=1pt}},
  scal/.style={draw, thick, dashed, dash pattern=on 3pt off 2pt},
  blob/.style={circle, fill=black, minimum size=5.2pt, inner sep=0pt},
  sm/.style={circle, draw, thick, fill=white, minimum size=4.4pt, inner sep=0pt},
  lbl/.style={font=\footnotesize, inner sep=1.2pt},
  dcap/.style={font=\footnotesize, align=center, text width=3.8cm, anchor=north, inner sep=0pt},
}
\DeclareUnicodeCharacter{2212}{-}

\newcommand{\Q}[2]{Q^{(#2)}_{#1}}
\newcommand{\QQ}[1]{Q_{#1}}

\newcommand{\hc}{\mathrm{h.c.}}
\newcommand{\lrD}{\overleftrightarrow{D}}

\newcommand{\FR}{\textsc{FeynRules}}
\newcommand{\UFO}{\textsc{UFO}}
\newcommand{\MG}{\textsc{MadGraph5\_aMC@NLO}}

\newcommand{\SMEFTsim}{\textsc{SMEFTsim}}
\newcommand{\code}[1]{\texttt{#1}}

\title{Automate the geometric SMEFT at dimension eight}
\author[a,b]{Beno\^it Assi,}
\author[c]{Adam Martin}
\affiliation[a]{Perimeter Institute for Theoretical Physics, Waterloo, ON, Canada}
\affiliation[b]{Department of Physics, University of Toronto and Vector Institute, Toronto, ON, Canada}
\affiliation[c]{Department of Physics, University of Notre Dame, South Bend, IN, 46556 USA}
\emailAdd{bassi@perimeterinstitute.ca}
\emailAdd{amarti41@nd.edu}

\abstract{We present the first complete dimension-eight implementation of the geometric Standard Model Effective Field Theory (geoSMEFT) as UFO models for event generators such as \MG. The models contain all 674 baryon-number-conserving operators, with minimal flavour violation for the fermionic operators, and are also provided in the Murphy basis. We include dimension-six contributions and electroweak input-parameter corrections consistently through order $1/\Lambda^4$, and provide dimension-eight squared contributions as a diagnostic of EFT truncation. We validate the implementation against independent calculations and Ward identities. To make these models practical for collider studies, we develop a standalone tool that identifies all operators contributing to a given process at tree level, including those entering through input-parameter shifts. It reports their expected energy scaling and generates a restriction card to reduce the computational cost of event generation. The same geometric construction allows us to calculate two- and three-point functions beyond dimension eight, to arbitrary order in $1/\Lambda^2$. The models and the tool are publicly available.}

\begin{document}
\maketitle
\allowdisplaybreaks[1]
\raggedbottom
\emergencystretch=3em
\renewcommand{\topfraction}{0.9}
\renewcommand{\bottomfraction}{0.7}
\renewcommand{\textfraction}{0.1}
\renewcommand{\floatpagefraction}{0.8}
\setcounter{topnumber}{3}
\setcounter{totalnumber}{4}

\section{Introduction}
\label{sec:intro}

The Standard Model Effective Field Theory (SMEFT) extends the Standard Model (SM) by higher-dimensional operators consistent with its gauge symmetries~\cite{Buchmuller:1985jz,Brivio:2017vri},
\begin{equation}
    \mathcal{L}_{\rm SMEFT}
    = \mathcal{L}_{\rm SM}
    + \sum_{d>4} \sum_j
    \frac{C_j^{(d)}}{\Lambda^{d-4}}\,Q_j^{(d)} .
    \label{eq:lsmeft}
\end{equation}
Here $C_j^{(d)}$ denotes the Wilson coefficient of the operator $Q_j^{(d)}$, and $\Lambda$ is the scale of new physics. At dimension six, established tools connect the Warsaw basis~\cite{Grzadkowski:2010es} to event generators through \FR~\cite{Christensen:2008py,Alloul:2013bka} and the \UFO\ format~\cite{Degrande:2011ua,Darme:2023jdn,Brivio:2017btx,Brivio:2020onw,Degrande:2020evl,Alwall:2014hca}. At order $1/\Lambda^4$, the squared dimension-six contributions enter alongside the interference of dimension-eight operators with the SM. Dimension-eight operators also provide independent constraints on new physics, including positivity bounds derived from fundamental properties of scattering amplitudes~\cite{Adams:2006sv,Remmen:2019cyz,Zhang:2018shp,Bi:2019phv,Remmen:2020vts,Zhang:2021eeo}. However, their large number has limited most phenomenological studies to selected operators, including studies of Higgs physics~\cite{Hays:2018zze,Corbett:2021cil,Martin:2021cvs,Martin:2023fad,Corbett:2023yhk,Assi:2024zap,Dawson:2021xei,Ellis:2023zim}, electroweak precision observables~\cite{Corbett:2021eux,Corbett:2023qtg}, Drell-Yan production~\cite{Alioli:2020kez,Boughezal:2021tih,Boughezal:2022nof,Kim:2022amu}, diboson production~\cite{Degrande:2023iob,Martin:2023tvi}, and quartic gauge interactions~\cite{Ellis:2018cos}.

The complete dimension-eight operator basis is already known. Li et al.\ constructed a basis from the correspondence between operators and on-shell amplitudes~\cite{Li:2020gnx}, while Murphy independently presented a basis in a notation similar to the Warsaw basis~\cite{Murphy:2020rsh}. Their operator counts agree with the Hilbert series calculations of Refs.~\cite{Henning:2015alf,Marinissen:2020jmb}, which extend the earlier dimension-eight analysis of Ref.~\cite{Lehman:2015coa}. Automated methods for constructing operator bases, matching, and renormalisation have also been developed~\cite{Criado:2019ugp,Fonseca:2017lem,Harlander:2023ozs,Chala:2021cgt,Chala:2021pll,DasBakshi:2022mwk,Boughezal:2024zqa,DasBakshi:2026ief,Fuentes-Martin:2022jrf}. Several event-generator implementations are available, including the bosonic dimension-eight sector in \textsc{SmeftFR}~\cite{Dedes:2023zws,Dedes:2019uzs}, models for anomalous quartic gauge couplings~\cite{Eboli:2006wa,Eboli:2016kko,Durieux:2024zrg}, and implementations for specific processes~\cite{Degrande:2013kka,ElFaham:2025fow}. More recently, a complete dimension-eight \UFO\ model in the $\{M_W,M_Z,G_F\}$ input scheme was presented using \textsc{Matchete}~\cite{Hourtz:2026ijm,Fuentes-Martin:2022jrf}. The challenge is therefore not simply to implement the operator basis, but to make the complete theory practical for collider calculations.

In this work, we provide the first complete dimension-eight implementation compatible with the geometric SMEFT (geoSMEFT)~\cite{Helset:2020yio}. Our \UFO\ models contain all 674 baryon-number-conserving operators, with the flavour structure of the fermionic operators given by minimal flavour violation (MFV). The default basis differs from our corrected implementation of the Murphy basis only in the $\psi^2H^2D^3$ class, which we replace with operators that do not contribute to vertices with fewer than four fields~\cite{Kim:2022amu,Corbett:2023yhk}. This preserves the geoSMEFT description of the two- and three-point functions. We also provide models in the Murphy basis, including the corrections discussed in \cref{sec:basis}. Both the $\{M_W,M_Z,G_F\}$ and $\{\alpha,M_Z,G_F\}$ input schemes are implemented~\cite{Brivio:2017bnu,Brivio:2021yjb}. The combined models include the dimension-six sector of \SMEFTsim~\cite{Brivio:2017btx,Brivio:2020onw} in its \code{U35} setup, which keeps the leading flavour structure of each operator in powers of the Yukawa couplings and allows complex Wilson coefficients. The electroweak input-parameter relations are derived through order $1/\Lambda^4$, including terms quadratic in the dimension-six coefficients~\cite{Hays:2020scx,Corbett:2021eux}.

Using the full basis in an event generator is computationally expensive, and identifying which operators contribute to a particular process is not straightforward. To address this, we develop a standalone program, the selector, which identifies all operators that can contribute to a given process at tree level. It includes operators that enter directly through interaction vertices and those that contribute indirectly through shifts in the electroweak input parameters. The selector also reports the expected energy dependence of each operator~\cite{Assi:2025zmp}, allowing users to identify potentially energy-enhanced contributions. It then automatically generates a restriction card for \MG, removing irrelevant operators during model import and reducing the work needed to prepare the model.

The models allow calculations consistently through order $1/\Lambda^4$, including dimension-six interference and squared contributions together with dimension-eight interference. The dimension-eight squared terms can also be calculated, providing a useful diagnostic of the EFT truncation, although they do not constitute a complete prediction at order $1/\Lambda^8$. We demonstrate these capabilities in four LHC studies, examining the energy dependence of different operator classes, contributions that do not interfere with the SM at tree level, and the relative size of dimension-eight interference and squared terms.

The geoSMEFT formulation also allows us to extend the implementation beyond dimension eight. We automate the operators contributing to two- and three-point functions at arbitrary order, with the corresponding input-parameter relations~\cite{Helset:2020yio,Hays:2020scx}. A model expanded through dimension $2N+4$ therefore includes these functions through order $1/\Lambda^{2N}$, allowing systematic higher-order predictions for electroweak and Higgs pole observables at tree level.

To construct the models, we write down the operators and generate their \FR\ expressions automatically, checking their gauge structure, Hermiticity, and CP properties. We validate the resulting models against independent implementations and through Ward identities.

The paper is organised as follows. \Cref{sec:geo} introduces the geoSMEFT and our choice of basis, while \cref{sec:basis} describes the operator conventions and classifications. \Cref{sec:pipeline} presents the model construction, input schemes, and selector, and \cref{sec:counting} describes the energy counting. \Cref{sec:validation} collects the validation tests, and \cref{sec:pheno} applies the models and the counting to four LHC studies. \Cref{sec:classes,sec:truncation} examine contributions that do not interfere with the SM at tree level and the size of dimension-eight squared terms relative to the interference. The appendices provide further conventions, validation details, the energy counting of four representative processes, and the complete results for each process (\cref{app:conventions,app:detail,app:lambda,app:studies}).

\section{The geometric SMEFT}
\label{sec:geo}

The geometric SMEFT (geoSMEFT) provides a systematic description of two- and three-point interactions to arbitrary order in the SMEFT expansion~\cite{Helset:2020yio}. We first review its construction, then explain how it guides our choice of dimension-eight basis.

The geometric SMEFT starts from the observation that the four real components $\phi^I$, $I=1,\ldots,4$, of the Higgs doublet can be treated as coordinates on a field space~\cite{Alonso:2015fsp,Alonso:2016oah}. A field redefinition $\phi\to\phi'(\phi)$ is then a change of coordinates, under which physical observables remain invariant. The Higgs kinetic term,
\begin{equation}
\frac12 h_{IJ}(\phi)(D_\mu\phi)^I(D^\mu\phi)^J,
\end{equation}
defines a metric $h_{IJ}$ on this space. In the SM the metric is flat, while higher-dimensional operators introduce curvature. The geometry therefore provides a description of the Higgs interactions that does not depend on the choice of field coordinates.

The geoSMEFT extends this description to gauge bosons and fermions by organizing the Lagrangian in terms of field-dependent functions~\cite{Helset:2018fgq,Helset:2020yio},
\begin{equation}
\begin{aligned}
\mathcal L \supset{}&
\tfrac12\, h_{IJ}(\phi)\,(D_\mu\phi)^I(D^\mu\phi)^J
-\tfrac14\, g_{AB}(\phi)\,\mathcal W^A_{\mu\nu}\mathcal W^{B,\mu\nu}
-V(\phi)
\\
&+ k^A_{IJ}(\phi)\,(D^\mu\phi)^I(D^\nu\phi)^J\,\mathcal W^A_{\mu\nu}
+ f_{ABC}(\phi)\,\mathcal W^A_{\mu\nu}\mathcal W^{B,\nu\rho}\mathcal W^{C,\mu}_{\rho}
\\
&+\big[Y(\phi)\,\bar\psi_1\psi_2
+d_A(\phi)\,\bar\psi_1\sigma^{\mu\nu}\psi_2\,\mathcal W^A_{\mu\nu}
+\hc\big]
+L_{I,A}(\phi)\,(D^\mu\phi)^I\,\bar\psi_1\gamma_\mu\sigma_A\psi_2
+\dots .
\end{aligned}
\label{eq:geo}
\end{equation}
Here $\mathcal W^A=\{W^1,W^2,W^3,B\}$, $\sigma_A=\{\sigma^I,\mathbf 1\}$, and flavour indices are suppressed. The omitted terms include gluonic interactions and CP-odd counterparts. The field-dependent functions, called connections, organize towers of higher-dimensional operators through their expansion in $H^\dagger H/\Lambda^2$. For example, the Higgs metric $h_{IJ}$ receives contributions from $Q_{H\Box}$ and $Q_{HD}$ at dimension six, and from $\Q{H^6}{1}$ and $\Q{H^6}{2}$ at dimension eight.

The advantage of this organization becomes clear when choosing an operator basis. Integration by parts and the equations of motion allow operators to be rearranged so that all contributions to vertices with at most three fields are contained in the connections~\cite{Helset:2020yio,Misiak:2018gvl,Kim:2022amu}. The two- and three-point functions are then determined entirely by these functions and their derivatives evaluated at the vacuum. They describe the masses, field normalizations, and interactions of the SM particles. The number of independent operator structures contributing to each connection reaches a fixed value at finite mass dimension. The connection $k^A_{IJ}$ acquires one additional CP-even structure at dimension ten~\cite{Helset:2020yio}, and the same structure with the dual field strength gives its CP-odd counterpart. The connections can therefore be constructed to arbitrary order in $v/\Lambda$~\cite{Helset:2020yio,Hays:2020scx}, even though the total number of SMEFT operators continues to grow. This simplification does not extend to vertices with four or more fields. The Warsaw basis already has the required structure at dimension six~\cite{Grzadkowski:2010es,Kim:2022amu,Corbett:2023yhk}.

The geoSMEFT has been used to derive background-field gauge fixing, Ward identities, and Feynman rules~\cite{Helset:2018fgq,Corbett:2019cwl,Corbett:2020ymv,Corbett:2020bqv}, and to calculate higher-order corrections to electroweak and Higgs observables~\cite{Corbett:2021eux,Corbett:2021cil,Martin:2021cvs,Martin:2023fad,Corbett:2023yhk,Assi:2024zap,Kim:2022amu,Martin:2023tvi}. It also provides a systematic treatment of input parameters, fermion masses, and renormalisation~\cite{Hays:2020scx,Trott:2023jrw,Talbert:2021iqn,Helset:2022pde,Assi:2023zid,Assi:2025fsm}, while the field-space geometry clarifies the relation between SMEFT and HEFT~\cite{Alonso:2016oah,Cohen:2020xca,Cohen:2021ucp,Alonso:2021rac,Gomez-Ambrosio:2022qsi,Gomez-Ambrosio:2022why,Salas-Bernardez:2022hqv}.

\section{The operator basis}
\label{sec:basis}

We construct our dimension-eight models from the Murphy basis~\cite{Murphy:2020rsh}. The following subsections establish the conventions and operator counting, describe the basis change required by the geoSMEFT, and present corrections to the published expressions and the CP assignments.

\subsection{Conventions and counting}
\label{sec:conventions}

We follow the conventions of the Murphy basis~\cite{Murphy:2020rsh}, with the covariant derivative and dual field strengths defined as
\begin{equation}
\begin{aligned}
D_\mu &= \partial_\mu + i g_1 \mathsf{y} B_\mu
+ i g_2 t^I W^I_\mu + i g_3 T^A G^A_\mu,\\
\widetilde X_{\mu\nu} &= \frac12\epsilon_{\mu\nu\rho\sigma}X^{\rho\sigma},
\qquad
\widetilde H_j = \epsilon_{jk}H^{\dagger k},
\end{aligned}
\end{equation}
where $t^I=\tau^I/2$ and $\epsilon_{0123}=+1$. We also use $iH^\dagger\lrD_\mu H=iH^\dagger(D_\mu H)-i(D_\mu H^\dagger)H$. Further conventions are collected in \cref{tab:conventions}. We group operators into 21 classes according to their field and derivative content, following the Murphy basis.

The Murphy basis contains 674 baryon-number-conserving operator structures, of which 438 have real coefficients and 236 have complex coefficients. Counting the real and imaginary parts of complex coefficients separately gives, in the one-generation limit,
\begin{equation}
N_{\rm term}=438+2\times236=910=895+15.
\end{equation}
Of these 910 terms, 15 vanish or become redundant for one generation, leaving 895 independent terms, in agreement with Hilbert series counting~\cite{Henning:2015alf}. Including the 98 independent baryon-number-violating terms gives the familiar total of 993. These counts refer to the operator basis, not the additional flavour structures introduced by MFV. \Cref{tab:catalogue-summary} summarizes the counting for all 21 operator classes.

We implement the dimension-eight basis assuming minimal flavour violation (MFV), with $U(3)^5$ flavour symmetry broken only by the SM Yukawa matrices. We retain independent flavour structures at leading order and with two additional Yukawa matrices, including independent four-fermion contractions. For four-fermion operators, we also retain structures with four additional quark Yukawa matrices that remain independent when only the top and bottom Yukawa couplings are nonzero. The leading structure keeps the original operator coefficient, while additional flavour structures have independent coefficients. Complex coefficients are retained. Of the 585 baryon-number-conserving fermionic operators, 584 have a nonzero structure within this expansion. Together, these fermionic operators give 2247 flavour structures and 3315 real parameters.

\begin{table}[t]
\centering
\small
\setlength{\tabcolsep}{5pt}
\renewcommand{\arraystretch}{1.12}
\begin{tabular}{@{}r l rr rrr rr@{}}
\toprule
 & & \multicolumn{2}{c}{Operators}
 & \multicolumn{3}{c}{Coefficient counting}
 & \multicolumn{2}{c}{CP parity} \\
\cmidrule(lr){3-4}
\cmidrule(lr){5-7}
\cmidrule(l){8-9}
 & Class
 & $\Delta B=0$ & $\Delta B\neq0$
 & Real & Complex & Components
 & Odd & Re odd \\
\midrule
 1 & $X^4$                & 43  & 0  & 43  & 0  & 43  & 17 & 0 \\
 2 & $H^8$                & 1   & 0  & 1   & 0  & 1   & 0  & 0 \\
 3 & $H^6D^2$             & 2   & 0  & 2   & 0  & 2   & 0  & 0 \\
 4 & $H^4D^4$             & 3   & 0  & 3   & 0  & 3   & 0  & 0 \\
 5 & $X^3H^2$             & 6   & 0  & 6   & 0  & 6   & 3  & 0 \\
 6 & $X^2H^4$             & 10  & 0  & 10  & 0  & 10  & 5  & 0 \\
 7 & $X^2H^2D^2$          & 18  & 0  & 18  & 0  & 18  & 7  & 0 \\
 8 & $XH^4D^2$            & 6   & 0  & 6   & 0  & 6   & 3  & 0 \\
 9 & $\psi^2X^2H$         & 48  & 0  & 0   & 48 & 96  & 0  & 25 \\
10 & $\psi^2XH^3$         & 11  & 0  & 0   & 11 & 22  & 0  & 0 \\
11 & $\psi^2H^2D^3$       & 15  & 0  & 14  & 1  & 16  & 0  & 0 \\
12 & $\psi^2H^5$          & 3   & 0  & 0   & 3  & 6   & 0  & 0 \\
13 & $\psi^2H^4D$         & 12  & 0  & 11  & 1  & 13  & 2  & 0 \\
14 & $\psi^2X^2D$         & 57  & 0  & 57  & 0  & 57  & 17 & 0 \\
15 & $\psi^2XH^2D$        & 86  & 0  & 80  & 6  & 92  & 40 & 3 \\
16 & $\psi^2XHD^2$        & 24  & 0  & 0   & 24 & 48  & 0  & 0 \\
17 & $\psi^2H^3D^2$       & 18  & 0  & 0   & 18 & 36  & 0  & 9 \\
18 & $\psi^4H^2$          & 55  & 10 & 34  & 31 & 96  & 2  & 0 \\
19 & $\psi^4X$            & 141 & 28 & 114 & 55 & 224 & 57 & 7 \\
20 & $\psi^4HD$           & 68  & 16 & 0   & 84 & 168 & 0  & 0 \\
21 & $\psi^4D^2$          & 47  & 6  & 39  & 14 & 67  & 0  & 0 \\
\midrule
 & \textbf{Total}
 & \textbf{674} & \textbf{60}
 & \textbf{438} & \textbf{296} & \textbf{1030}
 & \textbf{153} & \textbf{44} \\
\bottomrule
\end{tabular}
\caption{Dimension-eight operators organized into the 21 classes of
the Murphy basis. The real and complex columns distinguish the
coefficient types, while ``Components'' counts a complex coefficient
twice. The coefficient-counting columns include both
baryon-number-conserving and baryon-number-violating operators.
The CP columns count operators with CP-odd real coefficients
(``Odd'') and complex coefficients whose real parts are CP odd
(``Re odd''). These counts refer to the operator basis before
introducing the MFV flavour structures.}
\label{tab:catalogue-summary}
\end{table}

\subsection{Choice of dimension-eight basis}

To preserve the geoSMEFT description of two- and three-point functions, we modify one class of the Murphy basis, $\psi^2H^2D^3$. The fifteen operators in this class can generate momentum-dependent three-point interactions that are not contained in the geoSMEFT connections~\cite{Kim:2022amu,Corbett:2023yhk}. For example,
\begin{equation}
  \Q{l^2H^2D^3}{1}
  = i(\bar l_p\gamma^\mu D^\nu l_r)
  (D_{(\mu}D_{\nu)}H^\dagger H).
\end{equation}
Replacing the undifferentiated Higgs field by its vacuum expectation value produces a vertex involving two fermions and a gauge or Higgs boson. Such contributions are physically equivalent to operators with more fields through integration by parts and the equations of motion, but they prevent the geoSMEFT expressions for the three-point interactions from being applied directly.

We therefore replace these fifteen operators with a set in which every Higgs field carries a derivative, following Refs.~\cite{Kim:2022amu,Corbett:2023yhk} for the fourteen operators whose current contains a single fermion type. These operators have no vertices with fewer than four fields. For each fermion current $\psi=l,e,q,u,d$, we define
\begin{equation}
\begin{aligned}
  \Q{\psi^2H^2D^3}{g1}
  &= i(\bar\psi_p\gamma^\mu\psi_r)
  \big[
  (D^\nu H)^\dagger D_{(\mu}D_{\nu)}H
  - (D_{(\mu}D_{\nu)}H)^\dagger D^\nu H
  \big],\\
  \Q{\psi^2H^2D^3}{g2}
  &= (\bar\psi_p\gamma^\mu i\lrD_\nu\psi_r)
  \big[
  (D_\mu H)^\dagger D_\nu H
  + (D_\nu H)^\dagger D_\mu H
  \big],
\end{aligned}
\end{equation}
where $D_{(\mu}D_{\nu)}=(D_\mu D_\nu+D_\nu D_\mu)/2$. The corresponding triplet operators for $l$ and $q$ contain a $\tau^I$ in each bilinear. The remaining operator is
\begin{equation}
  \Q{udH^2D^3}{g}
  = i(\bar u_p\gamma^\mu d_r)
  \big[
  (D^\nu\widetilde H)^\dagger
  D_{(\mu}D_{\nu)}H
  \big] .
\end{equation}
It contributes $(C\,\Q{udH^2D^3}{g}+\hc)/\Lambda^4$ to the Lagrangian, with $C$ complex.
The fourteen operators with real coefficients are Hermitian as written and CP even. The real part of the coefficient of $\Q{udH^2D^3}{g}$ also multiplies the CP-even combination.

The two bases are related by integration by parts and the equations of motion. We check that they span the same space of contact amplitudes involving two fermions and two Higgs bosons, reproducing the Hilbert series counting for one, two and three generations~\cite{Henning:2015alf}. However, the full change of basis also shifts coefficients of operators with more fields and, through the Higgs equations of motion, dimension-six coefficients at order $m_H^2/\Lambda^4$. A direct translation of the fifteen coefficients is therefore insufficient to specify an equivalent point in the full theory. We consequently also provide models in the original Murphy basis.

\subsection{Hermiticity and corrections to the Murphy basis}

We modify 84 operator expressions in the published Murphy basis~\cite{Murphy:2020rsh}. Of these, 68 are anti-Hermitian as written, 14 are implemented through their Hermitian parts in the Murphy-basis models, and two vanish identically. We discuss these corrections below.

Of the 68 anti-Hermitian operators, 62 belong to the classes $\psi^2X^2D$ and $\psi^2XH^2D$. These contain $\lrD$ without the factor of $i$ required for Hermiticity. The remaining six belong to $XH^4D^2$, where the contraction of $D^\mu H^\dagger D^\nu H$, with or without $\tau^I$, with an antisymmetric field strength is anti-Hermitian. In each case, inserting a factor of $i$ makes the operator Hermitian. The missing factors in the $\psi^2XH^2D$ class were also noted in Ref.~\cite{DasBakshi:2026ief}.

The other 14 operators belong to $\psi^2H^2D^3$. Their Hermitian conjugates do not coincide with the printed expressions, and inserting a factor of $i$ does not make them Hermitian. In models using the original Murphy basis, we therefore implement these operators as
\begin{equation}
\frac12 C(Q+Q^\dagger),
\end{equation}
with $C$ real. In our default basis, these operators are replaced by the Hermitian operators introduced in the previous subsection.

Finally, as also noted in Ref.~\cite{DasBakshi:2026ief}, the operators $\Q{udWH^2}{1,2}$ vanish as printed in the Murphy basis. Their Higgs current $\widetilde H^\dagger\lrD{}_\mu^I H$ vanishes because it contracts two identical doublets with the symmetric matrix $\epsilon\tau^I$. However, the Hilbert series predicts two independent operators with this field content~\cite{Henning:2015alf}. We therefore use
\begin{equation}
\begin{aligned}
\Q{udWH^2}{1}
&= (\bar u_p\gamma^\nu d_r)
D^\mu(\widetilde H^\dagger\tau^I H)
W^I_{\mu\nu},\\
\Q{udWH^2}{2}
&= (\bar u_p\gamma^\nu d_r)
D^\mu(\widetilde H^\dagger\tau^I H)
\widetilde W^I_{\mu\nu},
\end{aligned}
\end{equation}
each supplemented by its Hermitian conjugate. Unlike the printed expressions, these operators are nonzero and independent modulo integration by parts and the equations of motion. We retain the original operator labels. In addition to these changes, we correct several typographical
errors in the published operator expressions, summarized in \cref{tab:errata}.
\begin{table}[t]
\centering
\small
\setlength{\tabcolsep}{6pt}
\renewcommand{\arraystretch}{1.2}
\begin{tabular}{@{}p{0.31\textwidth}p{0.61\textwidth}@{}}
\toprule
Operator(s) & Correction to the published expression \\
\midrule
Class 13: $\psi^2H^4D$
& Restore the barred fermion in each bilinear. \\

$\Q{leq^2HD}{6}$
& Replace $\gamma^\nu$ with $\gamma^\mu$ to match the derivative index. \\

$\QQ{lqd^2H^2}$
& Restore the charge-conjugation matrix in the $(lq)$ bilinear. \\

$\QQ{eq^2dH^2}$
& Restore the charge-conjugation matrix in the $(ed)$ bilinear. \\

$\Q{q^2H^2D^3}{4}$
& Correct the printed label from $D^2$ to $D^3$. \\

$\Q{lq^2uHD}{1,2,3}$
& Replace $\epsilon_{jn}H^{n\dagger}$ with the appropriate $\widetilde H$ contraction. \\
\bottomrule
\end{tabular}
\caption{Additional corrections to the operator expressions in the Murphy basis~\cite{Murphy:2020rsh}. The Hermiticity corrections and the two vanishing operators are discussed in the text.}
\label{tab:errata}
\end{table}

\subsection{CP parity}
\label{sec:cp}

We determine the CP parity of each dimension-eight operator from the transformation of its fields, using the rules in \cref{app:cp}. Under CP, an operator transforms as $Q\to\eta Q^\dagger$, where $\eta=\pm1$. For a complex coefficient, the real and imaginary parts multiply the Hermitian combinations $Q+Q^\dagger$ and $i(Q-Q^\dagger)$, with CP parities $\eta$ and $-\eta$, respectively.

A common shortcut is to classify an operator as CP odd when it contains an odd number of dual field strengths. This works for the dimension-six operators with real coefficients, but fails for 170 of the 438 corresponding dimension-eight operators. For example,
\begin{equation}
\Q{l^2q^2B}{1}
= (\bar l\gamma^\mu l)(\bar q\gamma^\nu q)B_{\mu\nu}
\end{equation}
is CP odd despite containing no dual field strength. Under CP, both fermion currents and the field strength change sign, giving an overall minus sign. Its partner $\Q{l^2q^2B}{2}$, containing $\widetilde B_{\mu\nu}$, is instead CP even. Similar effects occur in $\psi^2XH^2D$, where the Higgs current replaces one of the fermion currents, and in $\psi^2X^2D$, where the $SU(2)$ antisymmetric tensor can change the CP assignment. This classification also reproduces the known CP properties of operators contributing to neutral triple gauge couplings, which first arise at dimension eight~\cite{Degrande:2013kka}.

Of the 236 operators with complex coefficients, the real part is CP even for 192 operators and CP odd for the remaining 44. These assignments classify the operator structures before inserting the MFV flavour tensors. The Yukawa matrices can introduce additional CP violation through their complex phases.

\section{Implementation}
\label{sec:pipeline}

We describe the construction of the event-generator models, including the input-parameter treatment and higher-order extension, followed by the tools for selecting and running a process.

\subsection{Model construction}

We construct the models in \FR\ from the dimension-eight operator definitions of \cref{sec:basis}, together with the SM Lagrangian and, for the combined models, the dimension-six sector of \SMEFTsim~\cite{Brivio:2017btx,Brivio:2020onw}. Each operator is defined by its fields, derivatives and index contractions. We check the mass dimensions and gauge contractions of these definitions and construct Hermitian Lagrangian terms. We then implement the flavour structures, input parameter relations and field normalizations, and export the resulting vertices in \UFO\ format. Identically vanishing vertices are removed to avoid problems in \code{ALOHA}~\cite{deAquino:2011ub}. The 60 baryon-number-violating operators are included in the operator definitions but not exported, since their fermion-flow structures are not supported by \code{ALOHA}.

We provide eight model variants, corresponding to two dimension-eight bases, two electroweak input schemes and the choice of including dimension-six operators. The default models use the geoSMEFT-compatible basis of \cref{sec:basis}, while the alternative models use the corrected Murphy basis. In the default basis, \code{dim8\_mw} and \code{dim8\_is} contain dimension-eight operators, while \code{dim68\_mw} and \code{dim68\_is} also include dimension six. The suffixes denote the $\{M_W,M_Z,G_F\}$ and $\{\alpha,M_Z,G_F\}$ input schemes. The Murphy-basis variants carry the additional suffix \code{\_murphy}. In both schemes, we derive the input relations consistently through order $1/\Lambda^4$, including terms quadratic in dimension-six coefficients.

The dimension-eight fermionic interactions use the MFV expansion of \cref{sec:conventions}. By default, $Y_u$ and $Y_e$ are diagonal and $Y_d=V_{\rm CKM}\operatorname{diag}(y_d,y_s,y_b)$, with only the top and bottom Yukawa couplings nonzero. 

For the combined models, we use the dimension-six operators of \SMEFTsim~\cite{Brivio:2017btx,Brivio:2020onw} with its $U(3)^5$ flavour structure. The additional field and input-parameter corrections needed at order $1/\Lambda^4$ are described in \cref{sec:inputs}, and their coupling-order assignments in \cref{sec:select}.

\subsection{Input parameters}
\label{sec:inputs}

The electroweak parameters $g_1$, $g_2$ and $v$ are fixed using either $\{M_W,M_Z,G_F\}$ or $\{\alpha,M_Z,G_F\}$. In the SMEFT, higher-dimensional operators modify the relations between these inputs and the Lagrangian parameters through field normalizations, electroweak mixing and corrections to the measured processes. For example, operators contributing to muon decay shift the relation between $G_F$ and the Higgs vacuum expectation value, so the SMEFT value $\bar v_T$ differs from $\hat v_T=(\sqrt{2}G_F)^{-1/2}$. An operator can therefore affect a process through the input parameters even when it has no vertex in that process. We derive the input relations through order $1/\Lambda^4$, including dimension-six-squared terms, using the geoSMEFT connections~\cite{Helset:2020yio,Hays:2020scx,Corbett:2021eux,Corbett:2023yhk}. In the $\{M_W,M_Z,G_F\}$ scheme, the electromagnetic coupling receives SMEFT corrections, while in the $\{\alpha,M_Z,G_F\}$ scheme these corrections appear instead in the predicted $W$ mass.

We also fix $\alpha_s$, the fermion masses and the Higgs mass to their input values. The vacuum contributions of $Q_{HG}$ and $\Q{G^2H^4}{1}$ are absorbed into the gluon field and coupling, leaving interactions with at least one Higgs boson~\cite{Helset:2020yio,Trott:2023jrw}. Similarly, the mass contributions of $\psi^2H^3$ and $\psi^2H^5$ are absorbed into the Yukawa couplings, while the Higgs quartic is determined from the Higgs mass, including corrections to the potential and field normalization. All input and field corrections are expanded through order $1/\Lambda^4$. We also apply their first-order shifts to the dimension-six interactions, which generates additional terms at order $1/\Lambda^4$. These contributions carry the corresponding \MG\ coupling orders, $\mathrm{NP}=1$ and $\mathrm{NP}=2$.

In the $\{\alpha,M_Z,G_F\}$ scheme the $W$ mass is a prediction, so the Wilson coefficients shift it. The shifted mass enters the $W$ propagators and the phase space. \MG\ assigns EFT coupling orders only to couplings, so the mass shift carries no \code{NP} order and is present in every run, including the SM run \code{NP=0}. Following \SMEFTsim~\cite{Brivio:2020onw}, we keep the shifted mass in the propagators rather than expanding it. Selecting \code{NP\^{}2==1} or \code{NP\^{}2==2} therefore misses part of the correction at that order, and we recover it from the runs that contain the mass shift.

Let $C$ denote the Wilson coefficients of both sectors, with $C=0$ the SM, and define $t=1/\Lambda^2$. The mass shift enters the SM run and the dimension-six interference, and we expand their combination,
\begin{equation}
X(t)
=
\sigma_{\mathrm{NP}=0}(C)-\sigma_{\mathrm{NP}=0}(0)+\sigma_{\mathrm{NP}^2=1}(C)
=
x_1 t + x_2 t^2 + \mathcal{O}(t^3).
\label{eq:np0comb}
\end{equation}
The corrections through order $1/\Lambda^4$ are then
\begin{equation}
\sigma_{1/\Lambda^2}=x_1 t,
\qquad
\sigma_{1/\Lambda^4}=\sigma_{\mathrm{NP}^2=2}(C)+x_2 t^2+\mathcal{O}(t^3).
\end{equation}
The coefficient $x_1$ is the dimension-six interference together with the first-order change of the SM rate. The coefficient $x_2$ contains the change of the SM rate from the order-$1/\Lambda^4$ mass shift and from the square of the order-$1/\Lambda^2$ shift, and the change of the dimension-six interference from the order-$1/\Lambda^2$ shift. The term $\sigma_{\mathrm{NP}^2=2}$, which contains the dimension-six square, changes with the mass only at order $1/\Lambda^6$. In the dimension-eight-only models the mass first shifts at order $1/\Lambda^4$ and $\sigma_{\mathrm{NP}^2=1}$ vanishes, so $x_1=0$ and the missing term at this order is $\sigma_{\mathrm{NP}=0}(C)-\sigma_{\mathrm{NP}=0}(0)$.

We obtain $x_1$ and $x_2$ by evaluating $X(t)$ at several values of $\Lambda$, with the Wilson coefficients held fixed. Changing the fitted scales tests the extraction. The numerical validation is described in \cref{sec:validation}, and practical choices of fit scales are discussed in \cref{app:extraction}.

\subsection{Higher-order two- and three-point interactions}

The geoSMEFT makes higher-order calculations practical for electroweak and Higgs pole observables. At tree level, the $W$ and $Z$ partial widths and the decay $h\to f\bar f$ depend on two- and three-point interactions together with the input parameter relations. The geometric connections organize these interactions, allowing successive EFT corrections to be calculated within the implemented flavour expansion without constructing the complete operator basis at every higher dimension~\cite{Helset:2020yio}. Our models extend this sector through dimension twelve.

We implement this construction by generating the operator towers associated with the connections. Most are obtained from a dimension-eight operator $Q$ as
\begin{equation}
\frac{C_Q^{(8+2k)}}{\Lambda^{4+2k}}(H^\dagger H)^kQ,
\end{equation}
with an independent Wilson coefficient at each dimension. We also include the additional structure in $k^A_{IJ}$ that first appears at dimension ten and its CP-odd counterpart. Increasing the truncation order therefore extends a known set of towers, while retaining their flavour structures and gauge contractions.

We retain the Higgs dependence when generating the interactions, since both the vacuum values and the Higgs derivatives of the connections enter couplings such as $hVV$ and $h\bar f f$. In unitary gauge, before canonical field normalization,
\begin{equation}
(H^\dagger H)^k=\frac{(\bar v_T+h)^{2k}}{2^k}.
\end{equation}
We generate the full products $(H^\dagger H)^kQ$ in \FR, including the Higgs dependence of both the prefactor and $Q$. The two- and three-point classification counts the fields in the resulting interaction after symmetry breaking: $hVV$ and $h\bar f f$ are three-point vertices, whereas $hhVV$ is a four-point vertex.

The construction takes the desired order $N$ in $1/\Lambda^2$ as an input. Thus $N=2$ includes contributions through dimension eight, while $N=4$ extends the two- and three-point interactions through dimension twelve. We perform this extension in the $\{M_W,M_Z,G_F\}$ input scheme. Alongside the operator towers, we expand the field normalizations, electroweak mixing, input parameter relations, Higgs potential and Yukawa couplings through the same order. Products of lower-order corrections are retained whenever they contribute through $1/\Lambda^{2N}$. The determination of $G_F$ also includes the higher-dimensional four-lepton contributions to muon decay~\cite{Hays:2020scx}, to which we add the interference and squared terms from the Yukawa-suppressed scalar coupling.

We export the resulting interactions in \UFO\ format. Vertices with at most three fields keep every term through order $1/\Lambda^{2N}$, while vertices with four or more fields keep the terms through $1/\Lambda^4$, as in \code{dim68\_mw}. We provide the models with $N=2$ and $N=4$, \code{dim68\_mw\_geo2} and \code{dim68\_mw\_geo4}. Through order $1/\Lambda^4$ both are \code{dim68\_mw}, and in \code{dim68\_mw\_geo4} the higher orders are requested in \MG\ with \code{NP<=4}. We validate both models in \cref{sec:validation}.

\begin{figure}[!htbp]\centering
\includegraphics[width=0.90\textwidth,height=0.72\textheight,keepaspectratio]{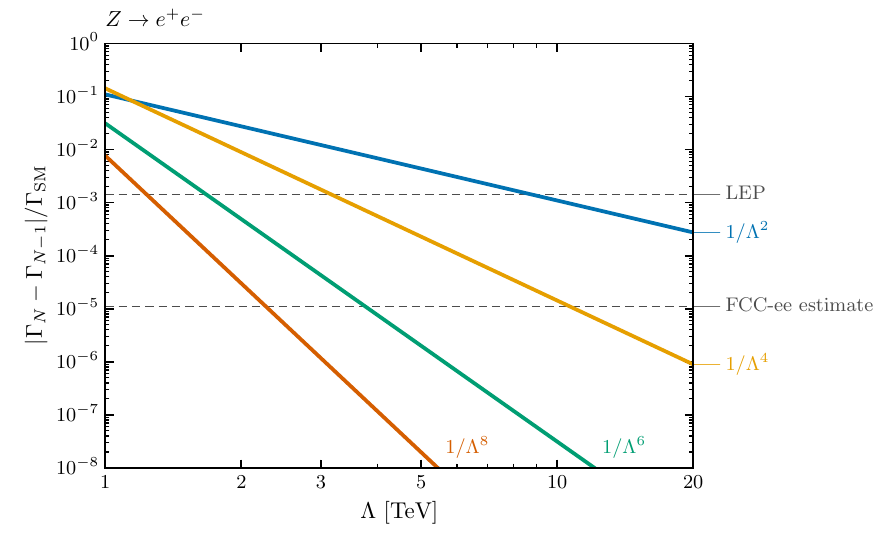}
\caption{Relative size of successive orders in $1/\Lambda^2$ in the leptonic $Z$ width, with $C_j^{(d)}=1$ for every Wilson coefficient that enters the decay. Each curve is $|\Gamma_N-\Gamma_{N-1}|/\Gamma_{\rm SM}$, where $\Gamma_N$ is the tree-level width in \code{dim68\_mw\_geo4} truncated at order $1/\Lambda^{2N}$ with every product of amplitudes at that order, so the curve at order $1/\Lambda^4$ includes the dimension-six terms at second order. The dashed lines are the relative uncertainty of the LEP measurement and the FCC-ee estimate.}
\label{fig:zwidth}
\end{figure}

\interfootnotelinepenalty=10000
We illustrate the higher-order calculation with the leptonic $Z$ width~\cite{Hays:2020scx,Corbett:2021eux}. \Cref{fig:zwidth} compares the size of each order in $1/\Lambda^2$ with the relative uncertainty of the LEP measurement~\cite{ParticleDataGroup:2024cfk} and an estimate for FCC-ee.\footnote{We obtain $1.1\times10^{-5}$ from the projected uncertainties on $\Gamma_Z$, $\sigma^0_{\rm had}$ and $R_\ell$~\cite{FCC:2025lpp}, using $\Gamma_{\ell\ell}=M_Z\Gamma_Z\sqrt{\sigma^0_{\rm had}/(12\pi R_\ell)}$ with lepton universality. We add the statistical and systematic uncertainties in quadrature and neglect the correlations between the measurements.}
For this benchmark, the correction at order $1/\Lambda^4$ remains larger than the estimated FCC-ee uncertainty for scales up to about 11~TeV. Computing the higher orders allows us to assess how far the EFT expansion must be retained for a given coefficient choice and experimental precision.

Extracting a pole width from a simple resonant process, such as $e^+e^-\to Z\to\mu^+\mu^-$, also requires controlling contributions that do not pass through the resonance. For a narrow, isolated resonance, two small parameters enter,
\begin{equation}
\epsilon=\frac{v^2}{\Lambda^2},\qquad
\delta=\frac{\Gamma}{M},
\label{eq:pole_expansion_parameters}
\end{equation}
where $M$ and $\Gamma$ are the resonance mass and width. For a continuum that varies smoothly across the resonance, the corrections relative to the leading pole amplitude $\mathcal A_0$ scale schematically as
\begin{equation}
\frac{\Delta\mathcal A_{\rm pole}^{(n)}}{\mathcal A_0}
\sim C_n\epsilon^n,\qquad
\frac{\Delta\mathcal A_{\rm continuum}^{(m)}}{\mathcal A_0}
\sim D_m\,\delta\,\epsilon^m,
\label{eq:pole_continuum_counting}
\end{equation}
where $C_n$ and $D_m$ represent the relevant coefficient combinations. The additional $\delta$ reflects the absence of the resonant enhancement~\cite{Corbett:2023yhk}. For comparable coefficients and coupling factors, an omitted continuum contribution can compete with a retained pole correction when $\epsilon^{n-m}\lesssim\delta$. Both expansions therefore help determine which terms must be included when using higher-order pole predictions to interpret scattering data.\footnote{For a simple $W/Z$ pole process with the complete tree-level prediction through dimension eight included, missing dimension-ten continuum contributions can enter at order $\delta\epsilon^3$. At $\Lambda=3$~TeV, $\epsilon\simeq0.007$ and $\delta\simeq0.03$ for the $W$ and $Z$~\cite{ParticleDataGroup:2024cfk}, so these can exceed dimension-twelve pole corrections of order $\epsilon^4$. This compares amplitude sizes. Their effects on measured rates also depend on interference.}

The relations between vertices with different Higgs multiplicities can help distinguish contributions from different operator dimensions. Comparisons across these channels must also include independent higher-point interactions. For example, a $q\bar qhZ$ contact interaction outside the geometric connections can contribute to $q\bar q\to hZ\to h\ell^+\ell^-$ with the same $Z$ resonance as the geometric terms~\cite{Corbett:2023yhk}. Unlike the continuum contributions of \cref{eq:pole_continuum_counting}, it is therefore not suppressed by $\delta$. Our complete dimension-eight models include these contributions. Beyond dimension eight, these analyses require both the higher-point vertices from the geometric towers and the independent higher-point interactions. As geometric descriptions of additional higher-point interactions are developed, they can be incorporated into the automation. This would extend the higher-order treatment to more general resonant processes.

\FloatBarrier
\subsection{Operator selection and code usage}
\label{sec:select}

Only a subset of the dimension-eight operators can contribute to a given process. We provide a selector that identifies possible tree-level contributions before diagram generation. Using the vertices generated by \FR, it connects operator interactions to the external particles through SM vertices and propagators, and includes contributions through the input parameter relations. This identifies operators that may contribute. It does not require their interference with the SM to be nonzero. Selection runs without \MG\ and requires no model rebuild.

The models and scripts are available at \url{https://github.com/benleo12/autom8smeft}. The released \UFO\ models can be imported directly into \MG\ without running \FR\ or Mathematica. \Cref{fig:pipeline} gives the command templates, with \code{PROCESS} standing for the requested process and \code{sel} naming the restriction card. A runnable example is given in \cref{sec:example}.

\definecolor{pipeA}{HTML}{2E5A88}   
\definecolor{pipeB}{HTML}{A8641B}   
\definecolor{pipeC}{HTML}{2F6B4F}   
\definecolor{pipeG}{HTML}{5F656E}   
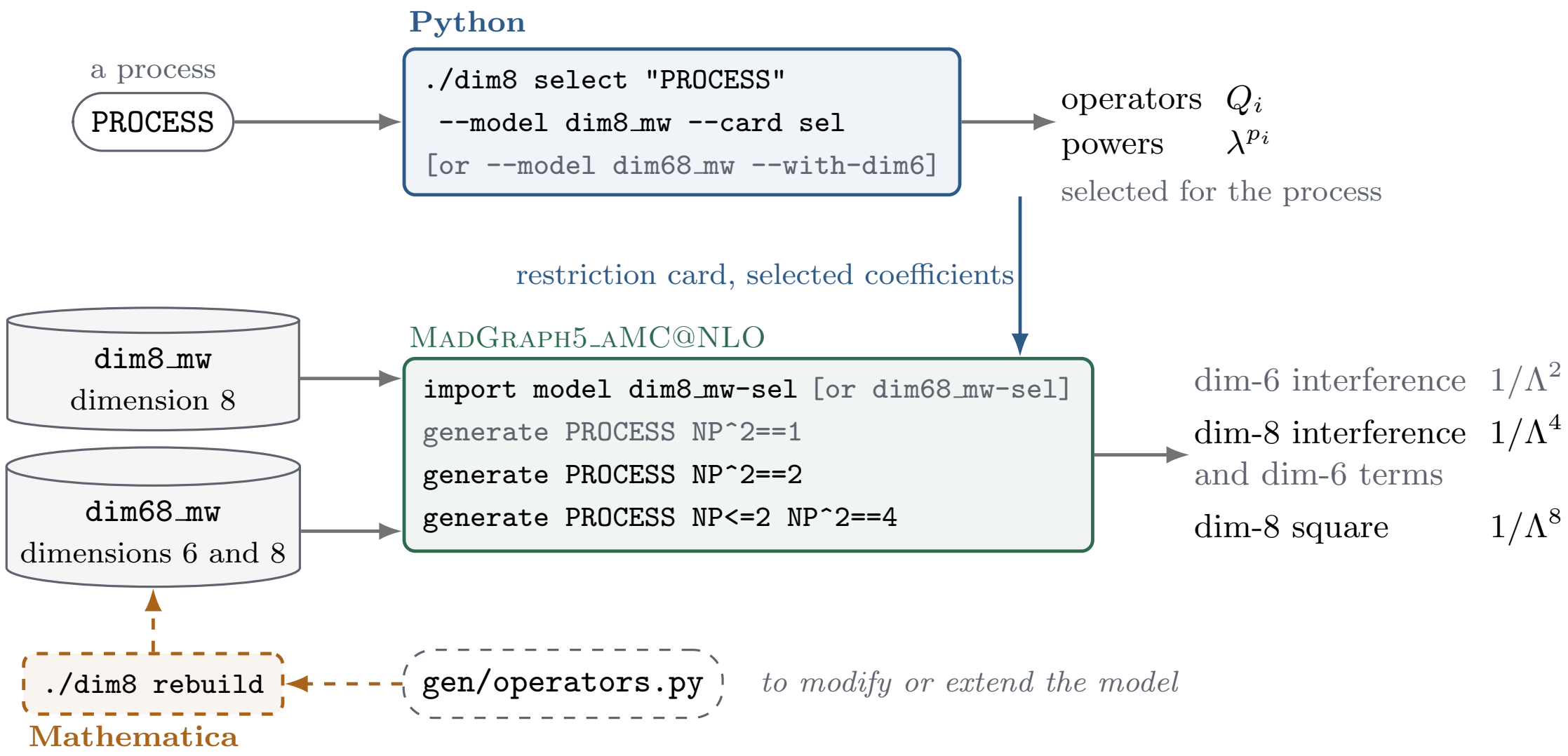
\begin{figure}[!htbp]
\centering
\begin{tikzpicture}[
  font=\footnotesize,
  tool/.style={draw=#1, line width=0.9pt, rounded corners=3pt, fill=#1!7, inner sep=5pt,
               align=left, font=\scriptsize},
  tag/.style={font=\scriptsize\bfseries, text=#1, inner sep=1.5pt},
  given/.style={draw=pipeG, line width=0.6pt, rounded corners=8pt, inner sep=5pt, fill=white},
  ret/.style={align=left, inner sep=1pt, anchor=west},
  note/.style={font=\scriptsize, text=pipeG, inner sep=1pt},
  flow/.style={-{Latex[length=2.2mm]}, line width=0.9pt, draw=black!55},
]

\node[given] (proc) at (0, 0) {\code{PROCESS}};
\node[note, anchor=south] at ([yshift=1pt]proc.north) {a process};
\tikzset{store/.style={cylinder, shape border rotate=90, aspect=0.13, draw=pipeG, line width=0.6pt, fill=black!4,
      minimum width=2.75cm, minimum height=1.08cm, align=center}}
\node[store] (model) at (0, -2.42)
  {\code{dim8\_mw}\\[-1pt]{\scriptsize dimension 8}};
\node[store] (model6) at (0, -3.86)
  {\code{dim68\_mw}\\[-1pt]{\scriptsize dimensions 6 and 8}};

\node[tool=pipeA, anchor=west] (sel) at (2.35, 0)
  {\code{./dim8 select "PROCESS"}\\[1pt]
   \code{\ --model dim8\_mw --card sel}\\[1pt]
   \textcolor{pipeG}{\code{[or --model dim68\_mw --with-dim6]}}};
\node[tag=pipeA, anchor=south west] at ([yshift=1pt]sel.north west) {Python};
\node[tool=pipeC, anchor=west] (mg) at (2.35, -3.14)
  {\code{import model dim8\_mw-sel} \textcolor{pipeG}{\code{[or dim68\_mw-sel]}}\\[1pt]
   \textcolor{pipeG}{\code{generate PROCESS NP\^{}2==1}}\\[1pt]
   \code{generate PROCESS NP\^{}2==2}\\[1pt]
   \code{generate PROCESS NP<=2 NP\^{}2==4}};
\node[tag=pipeC, anchor=south west] at ([yshift=1pt]mg.north west) {\mdseries\MG};

\node[ret] (ops) at ([xshift=9mm]sel.east)
  {\begin{tabular}{@{}l@{\hspace{2.2mm}}l@{}}
   operators & $Q_i$\\
   powers & $\lambda^{p_i}$
   \end{tabular}};
\node[note, anchor=north west, align=left] at ([yshift=-1pt]ops.south west)
  {selected for the process};
\node[ret] (xs) at ([xshift=9mm]mg.east)
  {\begin{tabular}{@{}l@{\hspace{2.2mm}}l@{}}
   \textcolor{pipeG}{dim-6 interference} & \textcolor{pipeG}{$1/\Lambda^2$}\\[2pt]
   dim-8 interference & $1/\Lambda^4$\\[-1pt]
   \textcolor{pipeG}{and dim-6 terms} & \\[2pt]
   dim-8 square & $1/\Lambda^8$
   \end{tabular}};

\node[tool=pipeB, dashed] (build) at (0, -5.3) {\code{./dim8 rebuild}};
\node[tag=pipeB, anchor=north west] at ([yshift=-1pt]build.south west) {Mathematica};
\node[given, dashed, anchor=west] (enc) at (2.35, -5.3) {\code{gen/operators.py}};
\node[note, anchor=west, font=\scriptsize\itshape] at ([xshift=3mm]enc.east) {to modify or extend the model};

\coordinate (card) at ([xshift=-7mm]mg.north east);
\draw[flow] (proc) -- (sel);
\draw[flow] (model.east) -- (mg.west |- model.east);
\draw[flow] (model6.east) -- (mg.west |- model6.east);
\draw[flow] (sel.east) -- (ops.west);
\draw[flow] (mg.east) -- (xs.west);
\draw[flow, draw=pipeA] (card |- sel.south) -- node[left, note, text=pipeA, align=right]
  {restriction card, selected coefficients} (card);
\draw[flow, dashed, draw=pipeB] (enc.west) -- (build.east);
\draw[flow, dashed, draw=pipeB] (build.north) -- (model6.south);

\end{tikzpicture}
\caption{Workflow for selecting operators and generating a process. The selector uses
  the model's vertices to find contributing operators and create a restriction
  card. MadGraph then computes the requested EFT contributions. Grey entries
  apply to the combined dimension-six/eight models. The dimension-eight square
  is obtained in a dimension-eight-only model, or with the dimension-six
  coefficients switched off in a combined model. Rebuilding the model with
  FeynRules requires Mathematica, but selecting and running operators does not.}
\label{fig:pipeline}
\end{figure}

For combined models, the selector also accounts for contributions with two dimension-six insertions, including corrections to dimension-six interactions from field and input parameter shifts. Choosing \code{--model dim68\_mw --with-dim6} retains the selected coefficients from both dimensions. Without \code{--with-dim6}, the written restriction card sets the dimension-six coefficients to zero.

The selector also uses the energy counting of \cref{sec:counting} to assign powers of $\lambda$ to the contributions. With \code{--max-power N}, the restriction card retains operators whose assigned power lies within $N$ powers of $\lambda$ above the SM amplitude. In combined models, it also retains dimension-six coefficients that enter only with a second insertion, for which the counting assigns no power. Other entries without an assigned power are excluded by this cut. This optional selection focuses the calculation on contributions expected to grow fastest relative to the SM at high energies, $v\ll E\ll\Lambda$.

Wilson coefficients are set in the run's parameter card: dimension-eight coefficients appear in \code{DIM8}, additional MFV flavour coefficients in \code{DIM8F}, and dimension-six coefficients in \code{DIM6}. The nonzero values in a restriction card are placeholders that preserve the selected parameters during model import. Their physical values are set subsequently in the parameter card. Inputs that vanish at import can also remove interactions. Calculations requiring nonzero light-fermion Yukawa couplings therefore use restriction cards generated with those couplings nonzero. Each model also provides cards for individual operator classes and useful combinations.

In \MG, the coupling order $\mathrm{NP}$ counts powers of $1/\Lambda^2$. Thus \code{NP\^{}2==1} selects the dimension-six interference, while \code{NP\^{}2==2} selects contributions at order $1/\Lambda^4$. These include the dimension-eight interference, the square of the first-order dimension-six amplitude, and the SM interference with amplitude terms quadratic in dimension-six coefficients. In a dimension-eight-only model, \code{NP\^{}2==2} gives the dimension-eight interference, and \code{NP<=2 NP\^{}2==4} gives its square. The additional treatment of the shifted $W$ mass in the $\{\alpha,M_Z,G_F\}$ scheme is described in \cref{sec:inputs}. \Cref{sec:example} gives the commands for a first calculation.

The selector identifies which operators matter for the requested process and writes a restriction card for \MG. \MG\ already finds the allowed diagrams, but it first prepares the model, including interactions irrelevant to that process. The restriction card removes those interactions during import, reducing this preparation work. The operator list and energy counting can also guide calculations with other event generators.

\section{Operator contributions and energy scaling}
\label{sec:counting}

The selector identifies operators that can contribute to a requested process. Here we explain the possible contributions through contact interactions, exchange diagrams and input-parameter shifts, and introduce a general counting of their energy dependence. This counting helps identify the interactions that can become most visible relative to the SM at high energies. We then describe how the rules apply to complete tree diagrams and different polarizations. Explicit amplitude expansions for representative processes are collected in \cref{app:lambda}.

\subsection{Routes into a process}
\label{sec:routes}

\begin{figure}[!htbp]
\centering
\begin{tikzpicture}
\begin{scope}[xshift=0cm]
  \node[blob] (v) at (0,0) {};
  \draw[ferm] (-1.15,0.78) -- (v); \draw[ferm] (v) -- (-1.15,-0.78);
  \draw[vec]  (v) -- (1.15,0.78); \draw[vec] (v) -- (1.15,-0.78);
  \node[lbl, anchor=east] at (-1.19,0.80) {$q$};
  \node[lbl, anchor=east] at (-1.19,-0.80) {$\bar q$};
  \node[lbl, anchor=west] at (1.19,0.80) {$W^+$};
  \node[lbl, anchor=west] at (1.19,-0.80) {$W^-$};
  \node[dcap] at (0,-1.35) {(a) $\psi^2X^2D$};
\end{scope}
\begin{scope}[xshift=6.4cm]
  \node[sm]   (a) at (-0.60,0) {};
  \node[blob] (b) at ( 0.60,0) {};
  \draw[ferm] (-1.70,0.78) -- (a); \draw[ferm] (a) -- (-1.70,-0.78);
  \draw[vec]  (a) -- (b);
  \draw[vec]  (b) -- (1.70,0.78); \draw[vec] (b) -- (1.70,-0.78);
  \node[lbl, anchor=east] at (-1.74,0.80) {$q$};
  \node[lbl, anchor=east] at (-1.74,-0.80) {$\bar q$};
  \node[lbl, anchor=west] at (1.74,0.80) {$W^+$};
  \node[lbl, anchor=west] at (1.74,-0.80) {$W^-$};
  \node[lbl] at (0,0.34) {$\gamma, Z$};
  \node[dcap] at (0,-1.35) {(b) $X^3H^2$};
\end{scope}
\begin{scope}[xshift=0cm, yshift=-3.3cm]
  \node[sm] (a) at (0,0.72) {};
  \node[sm] (b) at (0,-0.72) {};
  \draw[ferm] (-1.15,0.72) -- (a); \draw[ferm] (a) -- (b); \draw[ferm] (b) -- (-1.15,-0.72);
  \draw[vec]  (a) -- (1.15,0.72); \draw[vec] (b) -- (1.15,-0.72);
  \node[lbl, anchor=east] at (-1.19,0.74) {$q$};
  \node[lbl, anchor=east] at (-1.19,-0.74) {$\bar q$};
  \node[lbl, anchor=west] at (1.19,0.74) {$W^+$};
  \node[lbl, anchor=west] at (1.19,-0.74) {$W^-$};
  \node[lbl, anchor=west] at (0.10,0) {$q'$};
  \node[dcap] at (0,-1.35) {(c) $\Q{l^2H^4D}{2,4}$};
\end{scope}
\begin{scope}[xshift=6.4cm, yshift=-3.3cm]
  \node[blob] (v) at (0,0) {};
  \draw[glu]  (-1.15,0.78) -- (v); \draw[glu] (-1.15,-0.78) -- (v);
  \draw[vec]  (v) -- (1.15,0.78); \draw[vec] (v) -- (1.15,-0.78);
  \node[lbl, anchor=east] at (-1.19,0.80) {$g$};
  \node[lbl, anchor=east] at (-1.19,-0.80) {$g$};
  \node[lbl, anchor=west] at (1.19,0.80) {$W^+$};
  \node[lbl, anchor=west] at (1.19,-0.80) {$W^-$};
  \node[dcap] at (0,-1.35) {(d) $X^4$, $X^2H^2D^2$};
\end{scope}
\end{tikzpicture}
\caption{Four examples of dimension-eight contributions to $pp\to W^+W^-$. A filled circle is a dimension-eight vertex and an open circle an SM one.}
\label{fig:routes}
\end{figure}
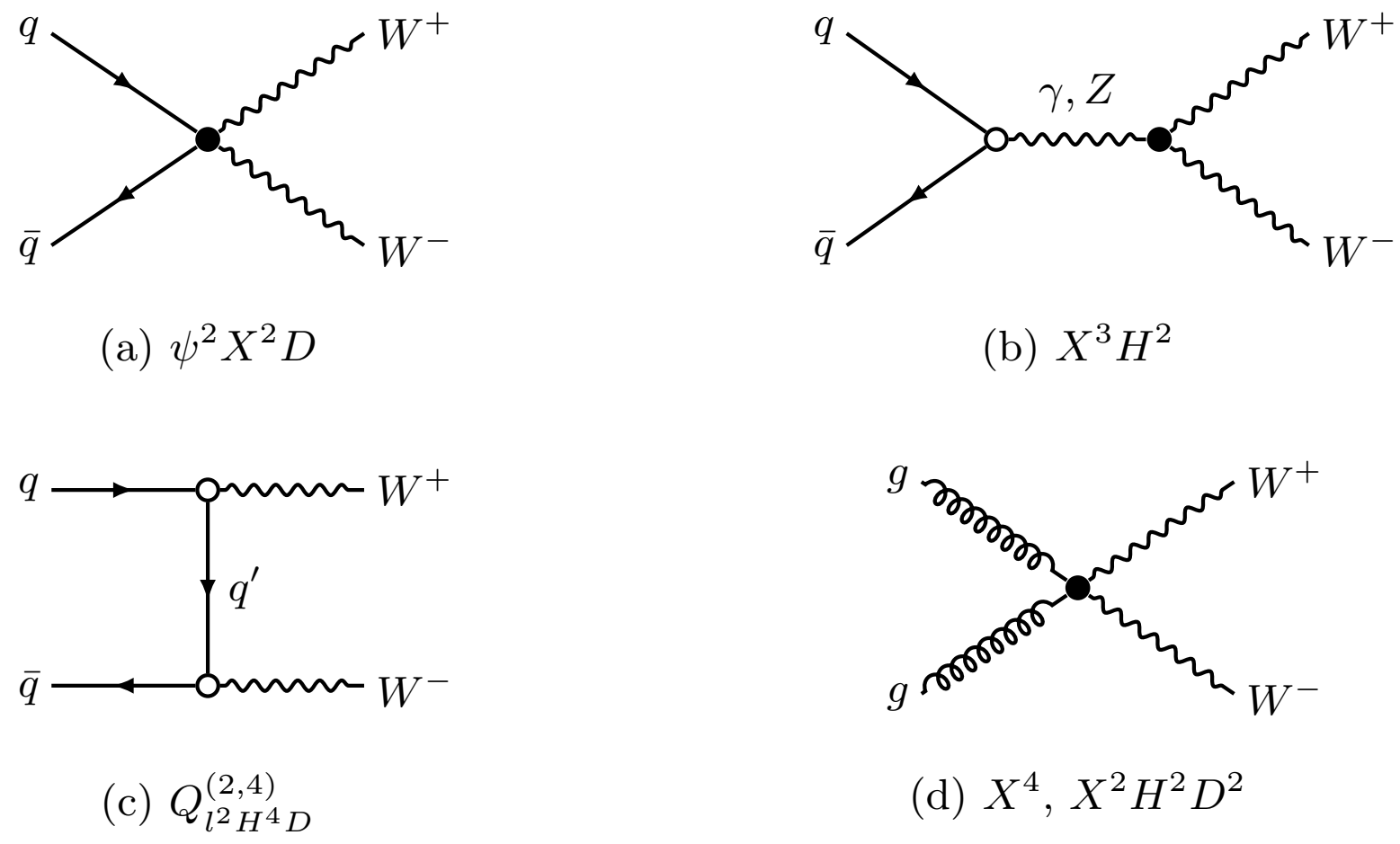

Operators can contribute through contact interactions involving all external particles, through exchange diagrams, or through shifts of the input parameters. \Cref{fig:routes} illustrates these possibilities in one process, $pp\to W^+W^-$. The selection procedure applies the same tests to any requested external state.

The selector takes a vertex from a SMEFT operator and tries to connect it to the particles of the process using SM vertices. If it can, the operator is kept. In \cref{fig:routes}(a), the new vertex already has all four particles of the process, so nothing else is needed. In \cref{fig:routes}(b), it has the two $W$ bosons and a $Z$. An SM vertex connects the $Z$ to the incoming quark pair, completing the diagram. 
The selector tries each vertex supplied by the operator and can add chains of SM vertices if needed. It checks whether these connections are possible without calculating matrix elements.\footnote{Results for each propagating field and group of external particles are stored and reused. Each SM vertex in the search has at least three legs, so each recursive step involves a smaller group of external particles. The search therefore terminates without an imposed depth limit.}

For two dimension-six insertions, the selector tests pairs of vertices from the same operator or from two different operators. It connects the pair through a propagator or a chain of SM interactions, then completes the remaining connections with SM vertices. Operators that can contribute this way through order $1/\Lambda^4$ are retained.

Input-parameter contributions are handled separately. The code checks whether the process admits an SM tree containing an electroweak vertex. If so, it also retains the operators entering the electroweak input relations of the selected model. Their coefficients can modify the electroweak parameters in the amplitude even when their own interaction vertices do not enter its tree diagrams. In \cref{fig:routes}, panel (c) illustrates an input-parameter shift. Panel (d) shows a contact interaction in a channel without a tree-level SM amplitude. The selector retains this contribution as well. With dimension-six coefficients set to zero, its first tree-level contribution to the partonic cross section is the dimension-eight square, at order $1/\Lambda^8$.

\subsection{Energy counting}

High-energy collider measurements offer sensitivity to heavy new particles through their virtual effects. Higher-dimensional interactions can generate contributions that grow relative to the SM prediction, making the tails of kinematic distributions promising places to search for deviations. Operators of the same mass dimension can differ substantially in this behaviour because they carry different powers of energy and the Higgs vacuum expectation value. We use the energy counting of Ref.~\cite{Assi:2025zmp} in the regime $v\ll E\ll\Lambda$ to identify contributions with the strongest growth relative to the SM. This organizes the calculation around the effects that can become most visible above the SM background, while allowing subleading contributions to be included systematically.

The high-energy expansion complements the geoSMEFT treatment of pole observables. In an on-shell two-body decay, such as $Z\to f\bar f$ or $h\to f\bar f$, the kinematics are fixed by the particle masses and the relevant expansion is in $v/\Lambda$~\cite{Helset:2020yio,Assi:2025zmp}. Scattering processes with four or more external particles can instead probe energies far above $v$. Three-point vertices can enter these processes through internal lines, as in panel (b) of \cref{fig:routes}, and their energy dependence must then be assessed within the complete diagram.

We organize the double expansion in $v/\Lambda$ and $E/\Lambda$ using a single dimensionless power counting parameter,
\[
(\Lambda,E,v)\sim(\lambda^{-3},\lambda^{-2},\lambda^{-1}).
\]
Consider a dimension-eight operator containing $N_f$ fermions, $N_H$ Higgs fields, $N_X$ field strengths and $N_D$ derivatives. For an $n$-leg contact interaction, its field content allows the leading scaling
\begin{equation}
\begin{aligned}
\mathcal A_n^{(8)} &\sim \frac{E^{8-n-p}v^p}{\Lambda^4}
\sim\lambda^{\,2n-4+p}, \\
p&=\max\!\left[N_f+N_X+N_H-n,\,0\right],
\end{aligned}
\label{eq:counting}
\end{equation}
where Wilson coefficients and SM coupling factors are suppressed. The integer $p$ is the minimum number of Higgs fields that must be replaced by their vacuum expectation value.

An unsuppressed SM amplitude has the reference scaling $\mathcal A_n^{\rm SM}\sim E^{4-n}$. The relative dimension-eight contribution therefore scales as
\[
\frac{\mathcal A_n^{(8)}}{\mathcal A_n^{\rm SM}}
\sim
\frac{E^4}{\Lambda^4}\left(\frac{v}{E}\right)^p.
\]
This expression makes the physical ordering transparent. Contributions with $p=0$ have the strongest relative growth, while each additional vacuum insertion introduces a suppression of $v/E$. At four legs, the corresponding powers are $\lambda^4$ for $E^4/\Lambda^4$, $\lambda^6$ for $E^2v^2/\Lambda^4$, and $\lambda^8$ for $v^4/\Lambda^4$. Their numerical importance also depends on the Wilson coefficients, couplings and interference with the SM.

A class generally contains several vertices with the same number of legs. Whether a particular operator and external state realize the leading power depends on the gauge contractions and polarization. For example, $\Q{G^2H^2D^2}{1}$ has a $gghh$ vertex with no vacuum insertion, giving $\lambda^4$. Its $ggW^+W^-$ vertex instead obtains both $W$ bosons from covariant derivatives acting on Higgs vacuum values. For transverse $W$ bosons, this contribution scales as $\lambda^6$.

The allowed vertex multiplicities follow from the same field content. Fermions and field strengths each supply at least one leg. A differentiated Higgs field also supplies a leg when replaced by its vacuum value, through the gauge-field term in $D_\mu\langle H\rangle$. At most, a non-Abelian field strength supplies two gauge bosons and each covariant derivative supplies one additional gauge boson. Thus
\begin{equation}
\max\!\left[N_f+N_X+N_{HD},\,3\right]
\leq n\leq
N_f+N_H+2N_X+N_D.
\label{eq:legs}
\end{equation}
Here $N_{HD}$ is the smallest number of differentiated Higgs fields among the terms of the operator or class under consideration, after writing derivatives as acting on individual fields. The lower bound of three excludes the vacuum energy, tadpoles and two-point terms, which are treated through the vacuum condition, field normalizations and input relations (\cref{sec:inputs}). Particular gauge contractions or external particles can further restrict the allowed vertices.

The differentiated Higgs fields raise the lower bound to four for $H^4D^4$, $X^2H^2D^2$, $\psi^2XH^2D$, $\psi^2XHD^2$ and $\psi^2H^3D^2$. These classes have no three-point vertex. The operators replacing $\psi^2H^2D^3$ in the default basis also have no three-point vertex (\cref{sec:basis}).

\begin{table}[tb]\centering
\begin{tabular*}{\textwidth}{@{\extracolsep{\fill}} r l ccc c r l ccc @{}}
\toprule
 & & \multicolumn{3}{c}{$\lambda$ at $n=$} & & & & \multicolumn{3}{c}{$\lambda$ at $n=$} \\
\cmidrule(lr){3-5}\cmidrule(lr){9-11}
 & class & 4 & 5 & 6 & & & class & 4 & 5 & 6 \\
\midrule
1 & $X^4$ & $\lambda^{4}$ & $\lambda^{6}$ & $\lambda^{8}$ & & 12 & $\psi^2H^5$ & $\lambda^{7}$ & $\lambda^{8}$ & $\lambda^{9}$ \\
2 & $H^8$ & $\lambda^{8}$ & $\lambda^{9}$ & $\lambda^{10}$ & & 13 & $\psi^2H^4D$ & $\lambda^{6}$ & $\lambda^{7}$ & $\lambda^{8}$ \\
3 & $H^6D^2$ & $\lambda^{6}$ & $\lambda^{7}$ & $\lambda^{8}$ & & 14 & $\psi^2X^2D$ & $\lambda^{4}$ & $\lambda^{6}$ & $\lambda^{8}$ \\
4 & $H^4D^4$ & $\lambda^{4}$ & $\lambda^{6}$ & $\lambda^{8}$ & & 15 & $\psi^2XH^2D$ & $\lambda^{5}$ & $\lambda^{6}$ & $\lambda^{8}$ \\
5 & $X^3H^2$ & $\lambda^{5}$ & $\lambda^{6}$ & $\lambda^{8}$ & & 16 & $\psi^2XHD^2$ & $\lambda^{4}$ & $\lambda^{6}$ & $\lambda^{8}$ \\
6 & $X^2H^4$ & $\lambda^{6}$ & $\lambda^{7}$ & $\lambda^{8}$ & & 17 & $\psi^2H^3D^2$ & $\lambda^{5}$ & $\lambda^{6}$ & $\lambda^{8}$ \\
7 & $X^2H^2D^2$ & $\lambda^{4}$ & $\lambda^{6}$ & $\lambda^{8}$ & & 18 & $\psi^4H^2$ & $\lambda^{6}$ & $\lambda^{7}$ & $\lambda^{8}$ \\
8 & $XH^4D^2$ & $\lambda^{5}$ & $\lambda^{6}$ & $\lambda^{8}$ & & 19 & $\psi^4X$ & -- & $\lambda^{6}$ & $\lambda^{8}$ \\
9 & $\psi^2X^2H$ & $\lambda^{5}$ & $\lambda^{6}$ & $\lambda^{8}$ & & 20 & $\psi^4HD$ & $\lambda^{5}$ & $\lambda^{6}$ & $\lambda^{8}$ \\
10 & $\psi^2XH^3$ & $\lambda^{6}$ & $\lambda^{7}$ & $\lambda^{8}$ & & 21 & $\psi^4D^2$ & $\lambda^{4}$ & $\lambda^{6}$ & $\lambda^{8}$ \\
11 & $\psi^2H^2D^3$ & $\lambda^{4}$ & $\lambda^{6}$ & $\lambda^{8}$ & &  & & & & \\
\bottomrule
\end{tabular*}
\caption{Leading powers allowed by the field content of each dimension-eight class at four, five and six legs. Within a column, a smaller power corresponds to stronger relative energy growth. Additional vacuum insertions give higher powers of $\lambda$. The entries $\lambda^4$, $\lambda^6$ and $\lambda^8$ at four, five and six legs have $p=0$ and scale as $E^4/\Lambda^4$ relative to an unsuppressed SM amplitude. A dash indicates the absence of a contact vertex with that multiplicity. The power realized in a process also depends on its external particles and polarizations.}
\label{tab:class-powers}
\end{table}

At four legs, seven classes allow $p=0$. Three are bosonic, $X^4$, $H^4D^4$ and $X^2H^2D^2$. Three contain a fermion pair, $\psi^2X^2D$, $\psi^2H^2D^3$ and $\psi^2XHD^2$. The seventh is $\psi^4D^2$. Additional external particles can expose stronger energy growth from other classes. Fourteen classes allow $p=0$ at five legs, and every class except $H^8$ and $\psi^2H^5$ allows it at six. The class $\psi^4X$ first contributes through a five-point vertex because none of its four fermions or its field strength can be replaced by a vacuum value.

Energy growth produces a leading interference contribution only when the SM has a nonzero amplitude with matching external helicities~\cite{Azatov:2016sqh}. With light-fermion masses and Yukawa couplings neglected, contributions whose external chiralities cannot match an SM amplitude do not interfere. This excludes the chirality-changing structures of $\psi^2XHD^2$ and eight of the 47 baryon-number-conserving operators in $\psi^4D^2$,
\[
\Q{leqdD^2}{1,2},\qquad
\Q{q^2udD^2}{1,2,3},\qquad
\Q{lequD^2}{1,2,3}.
\]
The counting therefore identifies the potential energy enhancement, while the available SM amplitudes determine whether that enhancement appears in the interference.

\subsection{Amplitude expansion and polarization}
\label{sec:lambda-amps}

Our earlier counting assigns powers of $\lambda$ to individual vertices~\cite{Assi:2025zmp}. Here the code builds tree diagrams for a chosen process and adds the powers from their vertices and propagators. The result gives the leading allowed powers of the amplitude $\mathcal A$, before squaring.

The counting rules of \cref{eq:counting} apply to each vertex and propagator. An SM three-point contribution scales as $\lambda^{-2}$ when a momentum or fermion wavefunctions supply a power of energy, and as $\lambda^{-1}$ when it is proportional to $v$. An SM four-point vertex has weight $\lambda^0$, and a propagator has weight $\lambda^4$. A dimension-$d$ vertex with $m$ legs and $p$ vacuum insertions has weight
\[
\lambda^{d-12+2m+p}.
\]
Adding these weights gives the power of each tree. For each operator class and topology, we retain the smallest power. 

The unsuppressed SM reference amplitude for $n$ external particles scales as $\lambda^{2n-8}$. A dimension-eight contribution four powers of $\lambda$ above this reference has relative growth $E^4/\Lambda^4$. Additional vacuum insertions reduce the energy enhancement as described in \cref{eq:counting}.

Knowing the diagram also tells us which legs are external. We can therefore specify their polarizations and count transverse and longitudinal $W/Z$ states separately.\footnote{Our event samples include all physical polarizations. Goldstone equivalence identifies the possible energy growth of the longitudinal contributions, which are already included in an unpolarized \MG\ calculation.} For each longitudinal boson, Goldstone equivalence~\cite{Chanowitz:1985hj,Wulzer:2013mza,Cuomo:2019siu} gives a Goldstone component and a residual gauge component carrying an extra factor $M_V/E\sim\lambda$. The code counts the allowed trees for both components using the same vertex and propagator rules, then keeps the smallest power.

The command \code{./dim8 series} constructs the expansion for a requested process using the selected model's vertices. The default combined model includes dimension-six and dimension-eight operators. The \code{--model} option selects the basis and input scheme, and the code checks that the vertex data agree with that model. Contributions entering only through the input relations are listed separately, with leading relative weights $\lambda^4$ at dimension six and $\lambda^8$ at dimension eight.

The restriction to one EFT insertion applies only to the displayed energy-counting series, which retain the leading power of each class in each topology. The combined models include the dimension-six square and amplitudes with two dimension-six insertions, as described in \cref{sec:select}. The powers classify the allowed amplitude contributions. The interference also depends on their helicities and on cancellations in the complete amplitude. \Cref{app:lambda} gives explicit expansions for the processes used in the phenomenological studies.

\section{Validation}
\label{sec:validation}

We validate the implementation through comparisons with independent models, numerical tests of Lorentz invariance and CP, and checks of the input-parameter treatment, operator selector and higher-order interactions.

We verify the geoSMEFT structure directly from the generated vertices, including Goldstone bosons. In the default basis, every dimension-eight operator contributing to a vertex with at most three fields belongs to the geoSMEFT connections. The corresponding check in the Murphy basis finds the fifteen additional operators of $\psi^2H^2D^3$.

For dimension eight, we compare our Feynman rules with \textsc{SmeftFR} v3~\cite{Dedes:2023zws} and models developed independently for our previous work. The \textsc{SmeftFR} comparison covers 87 of the 89 bosonic operators. For 44 operators, all Feynman rules of the difference between the two Lagrangians vanish. For the remaining 43 operators in $X^4$, we establish equality directly as polynomials in the field strengths. The two operators in $H^6D^2$ are excluded because the bases differ by equations of motion. Our earlier implementations provide additional operator-by-operator comparisons in both the bosonic and fermionic sectors. We also reproduce the dimension-eight interference in published studies of Higgs production in vector-boson fusion~\cite{Assi:2024zap}, Drell--Yan production~\cite{Boughezal:2021tih} and diboson production~\cite{Degrande:2023iob}, as well as the dimension-six and dimension-eight corrections to the $Z$-boson couplings up to order $1/\Lambda^4$~\cite{Corbett:2021eux}.

We compare the dimension-six sector with \SMEFTsim~\cite{Brivio:2017btx,Brivio:2020onw} in both input schemes. With dimension-eight coefficients set to zero, we vary all 85 real dimension-six parameters individually and compare the vertices at order $1/\Lambda^2$. All 1648 comparisons in the $\{M_W,M_Z,G_F\}$ scheme and all 1639 in the $\{\alpha,M_Z,G_F\}$ scheme agree. These comparisons include the input-parameter corrections and the operators $Q_{HG}$ and $\psi^2H^3$, whose implementation differs from \SMEFTsim. The comparison is restricted to first order because our models also include dimension-six-squared corrections.

We validate the fixed-order extraction described in \cref{sec:inputs} using the combined $\{\alpha,M_Z,G_F\}$ model, with all Wilson coefficients set to one. In $u\bar d\to e^+\nu_e$, the reconstructed $1/\Lambda^4$ contribution agrees with an independent expansion of the full squared amplitude at fixed phase-space points. This tests both the mass shift of the dimension-six interference and the contribution quadratic in the dimension-six mass shift. For the integrated cross section at 13.6~TeV, with $m_{e\nu}>100$~GeV and $\Lambda=3$~TeV, the full $1/\Lambda^4$ contribution is 26.1~fb. The \code{NP\^{}2==2} contribution evaluated at the unshifted mass supplies 14.6~fb. Including the extracted mass-shift terms reproduces the full result to within 0.2~fb.

The mass shifts can also change the sign of the leading correction. In $pp\to W^+W^-$ at 13.6~TeV with $\Lambda=3$~TeV, \code{NP\^{}2==1} gives $-1.38$~pb, while the complete $1/\Lambda^2$ contribution is $+0.53\pm0.04$~pb. At order $1/\Lambda^4$, the mass shifts add $+0.26\pm0.08$~pb to the 1.12~pb supplied by \code{NP\^{}2==2}. The quoted uncertainties are errors on the mean estimated from three independent random seeds.

We test Lorentz invariance in the exported \UFO\ models using the \code{check lorentz} command of \MG. For processes with external massless gauge bosons, this test is sensitive to violations of Ward identities. With all dimension-eight coefficients set to $0.7$ and $\Lambda=1$~TeV, all 29 tested processes pass in each of the four default-basis models. Omitting the SMEFT corrections to the input parameters and field normalizations causes eleven processes to fail, including $u\bar u\to gg$, $gg\to gg$, $gg\to ggg$ and $u\bar d\to W^+\gamma$. These controls demonstrate the importance of including those corrections consistently. With dimension-six coefficients switched on, all 29 processes pass at both \code{NP<=1} and \code{NP<=2} in the two combined models of the default basis. Electroweak processes also pass with both sectors switched on.

We test the CP assignments using five inclusive processes: $pp\to W^+W^-$, $ZZ$, $Zh$, $e^+e^-\gamma$ and $e^+e^-g$. For the CP-symmetric phase-space and luminosity combinations used here, CP-odd contributions cannot interfere with the SM in the total rate. We switch on the 153 CP-odd operators with real coefficients and the CP-odd components of the 236 operators with complex coefficients. Their interference is consistent with zero in all five processes, while their squared contributions remain nonzero. The additional flavour structures are tested with nonzero light-fermion masses and Cabibbo angle at $\Lambda=3$~TeV. Their 1108 CP-odd components also give interference consistent with zero, while the CP-even components give nonzero interference.

As a control, we repeat the tests with the dual-counting assignment of \cref{sec:cp}. The 86 operators classified as CP odd only by dual counting give nonzero interference, whereas the 84 classified as CP odd by our derivation give interference consistent with zero.

We further check the completeness of the operator selector against \MG. For representative processes, we generate diagrams using only the dimension-eight operators excluded by the selector and find no diagrams with a dimension-eight insertion. At dimension six, contributions from different vertices can cancel, so we compare squared matrix elements through order $1/\Lambda^4$, with and without the excluded coefficients. They agree at the tested phase-space points. These checks test the selection of contributions through both contact and exchange diagrams.

Finally, we test the higher-order construction in the $\{M_W,M_Z,G_F\}$ scheme against the geoSMEFT relations. The two-point functions and fermion masses must retain their canonical input values through order $1/\Lambda^{2N}$, with residual corrections scaling as $\Lambda^{-2(N+1)}$. We compare the left-handed $W$-electron coupling, at zero electron Yukawa coupling, with its all-orders expression~\cite{Helset:2020yio,Hays:2020scx} expanded to the chosen order. For vertices with at most three fields and no external Higgs boson, we compare the term linear in each tower coefficient with its dimension-eight parent. After factoring out the Wilson coefficients and powers of $\Lambda$, the ratio must be $(v^2/2)^k$ for the descendant $(H^\dagger H)^kQ$. At $N=2$, the construction also reproduces the two- and three-point vertices of the ordinary dimension-six/eight model.

We run these tests on the models with $N=2$ and $N=4$. In both, the residuals of the two-point functions, the fermion masses and the $W$-electron coupling fall as $\Lambda^{-2(N+1)}$. The comparison with a dimension-eight parent cannot test the two towers that have none. The first is the additional structure in $k^A_{IJ}$, with its CP-odd counterpart. At the vacuum its operator of dimension ten gives the vertices $W^+W^-\gamma$, $hZ\gamma$ and $hZZ$ of $\Q{WH^4D^2}{1}$, or of $\Q{WH^4D^2}{2}$ for the CP-odd counterpart, multiplied by $v^2/2$. The second is a four-lepton operator that first appears at dimension ten. It reaches the vertices only through the relation for $G_F$, where it acts as $\Q{l^4H^2}{1}$ multiplied by $-v^2/2$. We test both relations, and compare the members of dimension twelve of these towers with those of dimension ten.

We also check that every coupling of the exported models scales as $\Lambda^{-2n}$ with its \code{NP} order $n$, and that no term is dropped or counted twice, by comparing every vertex with \code{dim68\_mw} through $1/\Lambda^4$ and every vertex with at most three fields with the untruncated interactions beyond that order.

\section{Dimension-eight effects in four LHC studies}
\label{sec:pheno}

We use four LHC studies to illustrate how the operator contributions identified above appear in cross sections and kinematic distributions: diboson production, triboson production, Higgs-pair production in vector-boson fusion and high-mass Drell--Yan production. These twelve processes probe complementary combinations of gauge, Higgs and fermion interactions.

All studies use \code{dim8\_mw}, which contains no dimension-six sector. We calculate the interference of a single dimension-eight insertion with the SM at order $1/\Lambda^4$, including the input-parameter corrections. We also calculate the square of the single-insertion amplitude at order $1/\Lambda^8$ to compare its size with the interference. The calculations use \MG\ with the coupling orders and restriction cards described in \cref{sec:pipeline}. The study scripts are released with the models.

We consider proton-proton collisions at a centre-of-mass energy of $13.6$~TeV and take $\Lambda=3$~TeV. Each real Wilson coefficient switched on is set to one. For complex coefficients, both the real and imaginary parts are set to one. These choices also apply to the additional MFV flavour structures. At fixed coefficients and cuts, changing $\Lambda$ rescales the interference by $(3\,\mathrm{TeV}/\Lambda)^4$ and the squared contribution by $(3\,\mathrm{TeV}/\Lambda)^8$.

All cross sections are computed at leading order using the NNPDF2.3 QED LO set with $\alpha_s(M_Z)=0.130$~\cite{Ball:2013hta}. The initial-state partons and jets comprise gluons and the $u,d,s,c$ quarks and antiquarks. The renormalisation and factorisation scales are set event by event to $H_T/2$, and the gauge and Higgs bosons are on shell and undecayed. We use
\[
M_W=80.387~\mathrm{GeV},\qquad
M_Z=91.1876~\mathrm{GeV},\qquad
G_F=1.16637\times10^{-5}~\mathrm{GeV}^{-2},
\]
together with $m_h=125$~GeV and $m_t=172$~GeV. These electroweak inputs give $\alpha^{-1}=132.25$ at tree level.

The tables and figures use runs of $10^5$ events for each SM rate, interference and square, and of $10^6$ events where $10^5$ leaves a Monte Carlo error above five per cent. The kinematic cuts for all processes, the run records and the consistency checks are given in \cref{app:studies}.

\subsection{Cross sections and operator contributions}
\label{sec:results}

Switching on one operator class at a time separates the effects of different field contents, while retaining the interference among operators within that class in the squared term. A short worked example is provided in \cref{sec:example}.

We characterize each class by
\begin{equation}
\delta=\frac{\sigma_{1/\Lambda^4}}{\sigma_{\rm SM}},
\qquad
R=\frac{\sigma_{1/\Lambda^8}}{|\sigma_{1/\Lambda^4}|},
\label{eq:delta-r}
\end{equation}
where $\sigma_{1/\Lambda^4}$ is the interference with the SM and $\sigma_{1/\Lambda^8}$ denotes the square of the single-insertion dimension-eight amplitude. The signed quantity $\delta$ measures the change in the rate relative to the SM. The ratio $R$ compares the squared contribution with the magnitude of the interference.

\Cref{tab:master} collects one process from each study, and \cref{tab:grid} in \cref{app:studies} extends the interference results to all twelve processes. The pattern of nonzero entries identifies which interactions each process probes. Comparing their sizes then shows how strongly the chosen coefficients affect the inclusive rates. The following subsections discuss these patterns and the additional information in the kinematic distributions.
\begin{table}[tb]
\centering
\footnotesize
\setlength{\tabcolsep}{3.2pt}
\begin{tabular}{@{}r l r r r r r r r r@{}}
\toprule
 & & \multicolumn{2}{c}{$pp\to W^+W^-$} & \multicolumn{2}{c}{$pp\to W^+W^-W^+$} & \multicolumn{2}{c}{$pp\to HHjj$} & \multicolumn{2}{c}{$pp\to e^+e^-$} \\
\cmidrule(lr){3-4}\cmidrule(lr){5-6}\cmidrule(lr){7-8}\cmidrule(lr){9-10}
 & class & $\delta\,[10^{-4}]$ & $R$ & $\delta\,[10^{-4}]$ & $R$ & $\delta\,[10^{-4}]$ & $R$ & $\delta\,[10^{-4}]$ & $R$ \\
\midrule
\rowcolor{black!8}\textbf{1} & $\boldsymbol{X^4}$ & $\ast$ & $[81]$ & $+16.4$ & $1.0$ & -- & -- & -- & -- \\
2 & $H^8$ & -- & -- & -- & -- & $+5.77$ & $<0.01$ & -- & -- \\
3 & $H^6D^2$ & $-0.01$ & $<0.01$ & $-0.01$ & $<0.01$ & $-0.30$ & $<0.01$ & $-0.60$ & $<0.01$ \\
4 & $H^4D^4$ & -- & -- & $-0.03$ & $<0.01$ & $+11.4$ & $0.011$ & -- & -- \\
5 & $X^3H^2$ & $-0.01$ & $0.011$ & $-0.02$ & $0.27$ & -- & -- & -- & -- \\
6 & $X^2H^4$ & $-0.06$ & $0.12$ & $+0.44$ & $<0.01$ & $-13.3$ & $0.059$ & $-0.98$ & $<0.01$ \\
\rowcolor{black!8}\textbf{7} & $\boldsymbol{X^2H^2D^2}$ & $\ast$ & $[1.3]$ & $-0.44$ & $0.17$ & $+8.52$ & $6.4$ & -- & -- \\
8 & $XH^4D^2$ & $-0.01$ & $<0.01$ & $-0.09$ & $<0.01$ & $-0.54$ & $<0.01$ & -- & -- \\
9 & $\psi^2X^2H$ & -- & -- & -- & -- & -- & -- & -- & -- \\
10 & $\psi^2XH^3$ & -- & -- & -- & -- & -- & -- & -- & -- \\
11 & $\psi^2H^2D^3$ & $-0.13$ & $4.5$ & $-8.49$ & $2.4$ & $-1.10$ & $12$ & -- & -- \\
12 & $\psi^2H^5$ & -- & -- & -- & -- & -- & -- & -- & -- \\
13 & $\psi^2H^4D$ & $+0.19$ & $<0.01$ & $+2.12$ & $<0.01$ & $+16.5$ & $0.098$ & $-0.51$ & $<0.01$ \\
\rowcolor{black!8}\textbf{14} & $\boldsymbol{\psi^2X^2D}$ & $-4.30$ & $0.74$ & $-112$ & $5.2$ & -- & -- & $\ast$ & $[11]$ \\
15 & $\psi^2XH^2D$ & $+0.11$ & $0.41$ & $+2.65$ & $2.0$ & $+11.9$ & $4.1$ & -- & -- \\
16 & $\psi^2XHD^2$ & -- & -- & -- & -- & -- & -- & -- & -- \\
17 & $\psi^2H^3D^2$ & -- & -- & -- & -- & -- & -- & -- & -- \\
18 & $\psi^4H^2$ & $+0.45$ & $<0.01$ & $+0.68$ & $<0.01$ & $+22.5$ & $16$ & $+4.55$ & $<0.01$ \\
19 & $\psi^4X$ & -- & -- & -- & -- & -- & -- & -- & -- \\
20 & $\psi^4HD$ & -- & -- & -- & -- & -- & -- & -- & -- \\
\rowcolor{black!8}\textbf{21} & $\boldsymbol{\psi^4D^2}$ & -- & -- & -- & -- & -- & -- & $+20.8$ & $17$ \\
\bottomrule
\end{tabular}
\caption{How much each dimension-eight class changes four LHC processes at order $1/\Lambda^4$, and how large its square is compared with that change. All entries are for the LHC at $13.6$~TeV, with every coefficient of the class at one and $\Lambda = 3$~TeV, and the $HHjj$ and $e^+e^-$ samples carry the cuts of \cref{tab:grid}. $\delta$ is the interference with the SM (\code{NP\^{}2==2}) in units of $10^{-4}$ of the SM rate, and $R = \sigma_{1/\Lambda^8}/|\sigma_{1/\Lambda^4}|$ is the square of the class over its interference. The Monte Carlo error of every entry is below five per cent. An asterisk marks a nonzero squared contribution with no tree-level SM interference. The number in square brackets gives the square in units of $10^{-4}$ of the SM rate. A dash means no diagram. The shaded classes are discussed in the text.}
\label{tab:master}
\end{table}
The ratio depends on the coefficient choice and cuts. Under a common rescaling $C_i\to aC_i$ within a class, $R\to |a|R$. Its interpretation as a truncation diagnostic is discussed in \cref{sec:truncation}.

\subsection{Which classes contribute}
\label{sec:classes}

The processes probe complementary sets of interactions. The pattern of entries in \cref{tab:master,tab:grid} reflects the multiplicity and gauge structure of the operator vertices, together with the flavour assumptions of the benchmark.

Vertex multiplicity provides a first selection rule. The class $\psi^4X$ requires at least five external particles because none of its four fermions or its field strength can be replaced by a vacuum value. It therefore has no contribution to ordinary dilepton production, but becomes accessible when a photon is added to the final state. Additional final-state particles can thus expose interactions that are absent in a lower-multiplicity process.

The identities of the external particles matter as well. The four-point vertices of $X^4$, $H^4D^4$ and $X^2H^2D^2$ are purely bosonic. At tree level, they cannot connect to an initial quark pair while leaving only two outgoing bosons. They can, however, contribute to the interference in triboson production. Gauge contractions further distinguish the channels. The class $X^3H^2$ contributes to triboson channels containing charged $W$ bosons but gives no tree-level contribution to $ZZZ$, since three neutral gauge fields cannot saturate its antisymmetric $SU(2)$ contraction.

Six chirality-flipping classes are absent under the flavour and Yukawa restrictions used here:
\[
\psi^2X^2H,\quad \psi^2XH^3,\quad \psi^2H^5,\quad
\psi^2XHD^2,\quad \psi^2H^3D^2,\quad \psi^4HD.
\]
Their MFV coefficients contain Yukawa matrices. Only the top and bottom Yukawa couplings are retained in these benchmarks, and neither quark appears among the external particles. Restoring light-fermion Yukawa couplings makes these interactions available, with mass-suppressed interference in the chirality-preserving SM channels.

The asterisks identify nonzero squared contributions with no tree-level SM interference. In gluon-initiated diboson production, for example, $X^4$ and $X^2H^2D^2$ supply contact interactions, while the SM amplitude is loop induced. Their first contribution in our tree-level, dimension-eight-only calculation is therefore the squared amplitude at order $1/\Lambda^8$. These channels provide access to interactions absent from the quark-initiated interference. The size and energy dependence of their squared contributions are discussed in \cref{sec:truncation}.

\begin{figure}[!t]\centering
\includegraphics[width=0.95\textwidth,height=0.72\textheight,keepaspectratio]{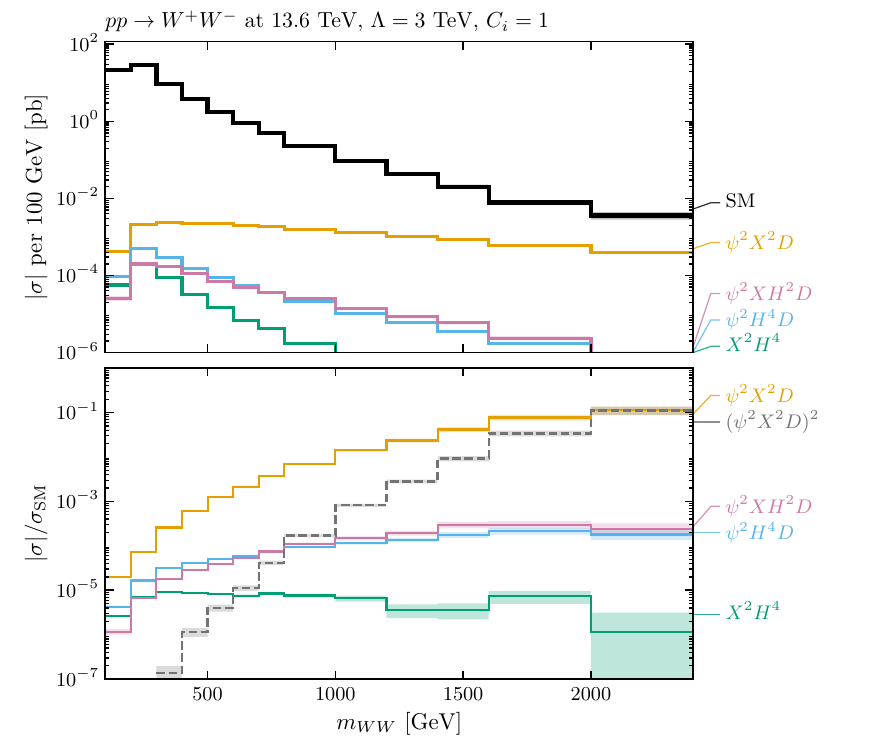}
\caption{The interference of $\psi^2X^2D$ in $pp \to W^+W^-$ grows with the mass of the boson pair to several per cent of the SM, and its square reaches the interference at the high-mass end. The upper panel shows the SM and the interference of the four classes named beside the curves, each switched on alone, and the lower panel the ratio of each interference to the SM. Both panels show absolute values, since the sign varies from class to class (\cref{tab:master}). The grey dashed curve is the square of $\psi^2X^2D$ over the SM. The shaded bands are the Monte Carlo errors. The cross section is given per $100$~GeV, with bins that widen with the mass, at $13.6$~TeV, $\Lambda = 3$~TeV and unit coefficients.}
\label{fig:ww}
\end{figure}

\subsection{Energy growth and the truncation}
\label{sec:truncation}

High-energy distributions can make heavy new physics more visible by enhancing the SMEFT contribution relative to the SM background. The same growth can make a prediction based on the interference sensitive to the terms it omits. We therefore examine both the interference relative to the SM and the size of the squared dimension-eight amplitude across invariant-mass bins.

Comparisons with calculable terms beyond the retained order have been used to assess truncation effects in dimension-six analyses~\cite{Hays:2018zze,Corbett:2021eux,Trott:2021vqa,Boughezal:2021tih,Kim:2022amu,Assi:2026vjx}. The diagnostic used here omits dimension-ten interference at order $1/\Lambda^6$ and the other terms required for a complete prediction at order $1/\Lambda^8$. When $R>1$, the calculated square exceeds the magnitude of the retained interference. This does not by itself determine whether the EFT expansion converges, since suppression or cancellation of the interference can also make $R$ large~\cite{Brivio:2022pyi}. All results below use $\Lambda=3$~TeV and unit coefficients. The distributions and cumulative upper cuts carry Monte Carlo errors obtained from the event weights and the cross-section normalisation, as described in \cref{sec:example}.

The gain in sensitivity is illustrated by $pp\to W^+W^-$ in \cref{fig:ww}. For the leading interference class, $\psi^2X^2D$, the magnitude of the interference rises from a few times $10^{-5}$ of the SM rate at low pair mass to several per cent above $1.4$~TeV. The class $\psi^2H^4D$, whose longitudinal contribution enters at $\lambda^6$, grows by a factor of about six between $200$~GeV and $1$~TeV, reaching $10^{-4}$ of the SM rate. Over the same range, the relative contribution of $X^2H^4$ stays nearly constant, between $0.7$ and $0.9\times10^{-5}$.

The square of $\psi^2X^2D$ grows more rapidly than its interference. It reaches 44 per cent of the interference magnitude between $1.6$ and $2$~TeV and overtakes it near $2.2$~TeV. The inclusive ratio, $R=0.74$, hides this variation. About 70 per cent of the square comes from $m_{WW}>3$~TeV, above the chosen value of $\Lambda$, compared with only 4 per cent of the interference. Restricting the sample to $m_{WW}<3$~TeV reduces $R$ to $0.24$. For unit coefficients at $\Lambda=3$~TeV, the expansion is therefore not under control at the highest pair masses. These results motivate examining the comparison in mass bins and under upper cuts, or at smaller coefficients, before using an interference-only prediction~\cite{Contino:2016jqw}.

\begin{figure}[tb]\centering
\includegraphics[width=0.95\textwidth,height=0.72\textheight,keepaspectratio]{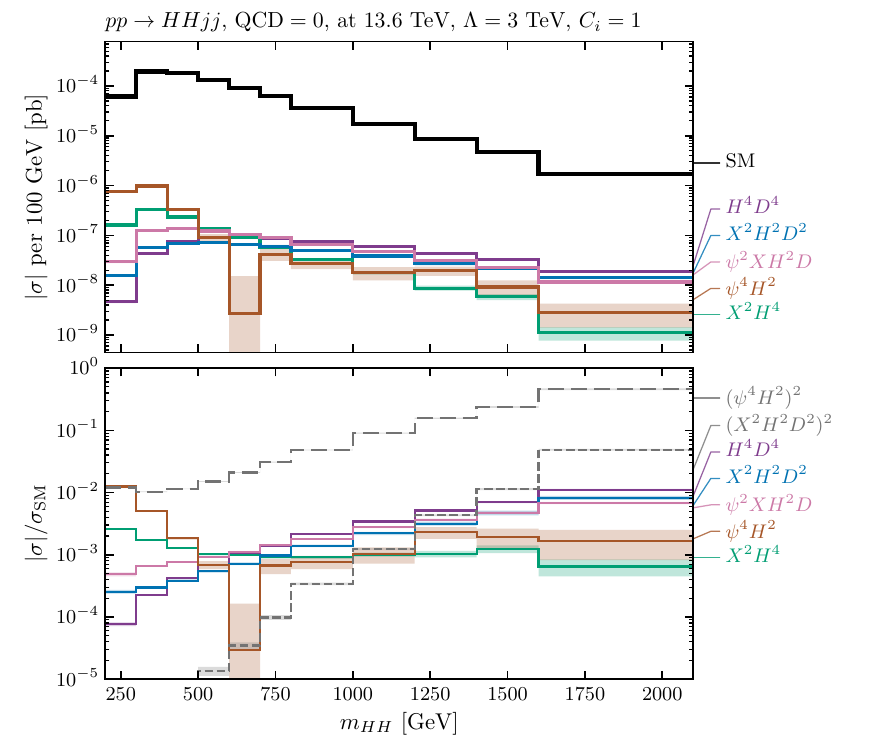}
\caption{Selected dimension-eight contributions to electroweak $HHjj$ production. The panels are as in \cref{fig:ww}, with the cuts of \cref{app:studies}, at $13.6$~TeV, $\Lambda = 3$~TeV and unit coefficients. The grey curves show the squared contributions relative to the SM, with short dashes for $X^2H^2D^2$ and long dashes for $\psi^4H^2$. The shaded bands are the Monte Carlo errors.}
\label{fig:vbfhh}
\end{figure}

The gluonic $X^4$ interactions illustrate a contribution that such a prediction misses entirely at tree level. Their square is $8.1\times10^{-3}$ of the SM rate, about twenty-two times the magnitude of the total dimension-eight interference in this benchmark. With the same coefficients, an upper cut near $1$~TeV is needed to keep the cumulative $X^4$ square below the magnitude of the $\psi^2X^2D$ interference. This is consistent with Refs.~\cite{Gillies:2024mqp,Gillies:2026tpc}, which find in gluon-fusion $WW$ production that the square of the gluonic $X^4$ interactions dominates at high mass. This comparison depends on the relative coefficients of the two classes and does not define a universal cutoff.

\Cref{fig:vbfhh} shows how the pattern changes in $pp\to HHjj$. Between $200$~GeV and $2$~TeV, the interference relative to the SM grows by about two orders of magnitude for $H^4D^4$ and by a factor of about thirty for $X^2H^2D^2$, reaching around one per cent in both cases. For $H^4D^4$, this growth arises in the longitudinal-vector-boson channel, as described in \cref{app:lambda}. By contrast, the magnitude of the $X^2H^4$ interference falls from $2.6\times10^{-3}$ of the SM rate at low pair mass to about $10^{-3}$ above $500$~GeV. The square of $X^2H^2D^2$ overtakes its interference near $1.2$~TeV, and its inclusive ratio is $R=6.4$.
For $\psi^4H^2$, the square becomes comparable to the interference near the Higgs-pair threshold, making an interference-only prediction unreliable over most of the plotted range. The dip in the interference reflects a change of sign.

\section{Conclusions}
\label{sec:conclusions}

We have automated the construction of tree-level SMEFT models through dimension eight and their use in collider calculations. The models include all 674 baryon-number-conserving dimension-eight operators, with minimal flavour violation for the fermionic interactions. We provide two operator bases and two electroweak input schemes, each with or without the dimension-six sector of \SMEFTsim. The combined models include the input and field corrections required for predictions through order $1/\Lambda^4$, including terms quadratic in the dimension-six coefficients. The extraction follows the scheme-dependent treatment in \cref{sec:inputs}, including the effects of the predicted $W$ mass in the $\{\alpha,M_Z,G_F\}$ scheme.

The default basis is adapted to the geoSMEFT, so that its two- and three-point interactions are organized by the geometric connections. Beyond dimension eight, we generate the associated operator towers and expand the field normalizations and input relations to the same order in the $\{M_W,M_Z,G_F\}$ scheme. This makes it possible to study successive EFT corrections to electroweak and Higgs pole observables without constructing the complete operator basis at every higher dimension. The extension retains independent coefficients at each dimension. Extracting pole observables from scattering data also requires assessing the narrow-width expansion and the contributions of higher-point interactions.

The operator selector makes the full dimension-eight basis practical for a specified process. It identifies possible tree-level contributions, including effects through the input parameters and contributions with two dimension-six insertions, and writes a restriction card for \MG. The resulting calculation uses the standard event-generator workflow without rebuilding the model. Energy counting then identifies contributions that can grow relative to the SM background, helping to focus searches on interactions and kinematic regions with enhanced sensitivity to heavy new physics.

Our four LHC studies cover twelve processes in diboson and triboson production, Higgs-pair production through vector-boson fusion, and high-mass Drell--Yan. They illustrate how different final states probe complementary interactions and how distributions reveal effects that inclusive rates can obscure. Comparing the dimension-eight interference with the square of its amplitude identifies benchmarks and kinematic regions where an interference-only prediction omits a sizeable calculated contribution. The comparison depends on the coefficients and cuts, and assessing EFT convergence also requires the other higher-order terms.

The combined models are particularly useful for estimating truncation uncertainties. Ref.~\cite{Assi:2026vjx} uses the square of the dimension-six amplitude, which dimension-six tools already compute, to estimate the complete order-$1/\Lambda^4$ dependence of dimension-six analyses. Our combined models give the complete order $1/\Lambda^4$, including the dimension-eight interference, in the same \MG\ calculation, so this estimate can now be tested directly.

The validation combines comparisons with independent implementations, vertex-level checks against \SMEFTsim, and numerical tests of the exported models. We release both the \UFO\ models and the code that builds and checks them. The models can be imported directly into \MG, and the worked example in \cref{sec:example} gives the steps for a first calculation. Retaining explicit fermion generation indices also provides a starting point for future extensions beyond minimal flavour violation.

\section*{Code availability}

The \UFO\ models, the operator selector and the code that builds and checks the models are publicly available at \url{https://github.com/benleo12/autom8smeft}.

\acknowledgments

We used Claude Fable 5.1 (Anthropic) to extend our private implementations of about 200 operators to the full dimension-eight operator set in both the Murphy and geoSMEFT-compatible bases, and to streamline the code that builds and checks the models.

Research at Perimeter Institute is supported in part by the Government of Canada through the Department of Innovation, Science and Economic Development and by the Province of Ontario through the Ministry of Colleges and Universities. 
B.A. was supported in part by
the Canadian AI Safety Institute Research Program at
CIFAR. This work was performed in part at the Aspen Center for Physics, with support for B.A. by a
grant from the Simons Foundation (1161654, Troyer).
The work of AM is partially supported by the National Science Foundation under Grant Number PHY-2412701. 

\clearpage
\appendix
\addtocontents{toc}{\protect\setcounter{tocdepth}{1}}
\section{Conventions and CP assignment}
\label{app:conventions}

\subsection{Field and operator conventions}

\Cref{tab:conventions} summarizes the field and operator conventions used in the implementation. These fix the signs and normalizations needed to compare the generated interactions with the operator definitions. The accompanying documentation records the definitions and checks supporting these choices.

\begin{table}[!htbp]
\centering
\small
\begin{tabular}{l l l}
\toprule
object & Murphy basis & our \FR\ form \\
\midrule
$SU(2)_w$, $SU(3)_c$ generators & $t^I = \tau^I/2$, $T^A = \lambda^A/2$ & $\tau^I \to 2\,\code{Ta}$, \code{T} \\
$d^{ABC}$, $\epsilon_{jk}$ & $\{T^A,T^B\}$ symbol, $\epsilon_{12}=+1$ & \code{dSUN}, \code{Eps} \\
$\widetilde H$ & $\widetilde H_j = \epsilon_{jk} H^{\dagger k}$ & $\code{Eps}\,\code{Phibar}$ \\
Levi-Civita, dual tensor & $\epsilon_{0123}=+1$, $\widetilde X = \tfrac12 \epsilon X$ & \code{Eps} with verified sign \\
$\sigma^{\mu\nu}$ & $\tfrac{i}{2}[\gamma^\mu,\gamma^\nu]$ & antisymmetrised \code{Ga} product \\
Hermitian derivative & $iH^\dag\lrD_\mu H$ & expanded two-term form \\
Wilson coefficients & $C\,Q/\Lambda^4$, $[C]=0$ & $c/\Lambda^4$, $\Lambda$ external \\
\bottomrule
\end{tabular}
\caption{Sign and normalization conventions used in the operator implementation. The accompanying documentation gives the full dictionary and the evidence supporting each choice.}
\label{tab:conventions}
\end{table}

We use the covariant derivative
\begin{equation}
D_\mu=\partial_\mu+igA_\mu,
\end{equation}
implemented in \FR\ by setting \code{FR\$DSign = -1}. The \FR\ default uses the opposite sign. The two conventions are related by changing the sign in the covariant derivative together with $A_\mu\to-A_\mu$. Under this simultaneous change, the covariant derivative and SM predictions are unchanged, while each field strength changes sign. The Wilson coefficients therefore satisfy
\begin{equation}
C_i^{\rm default}=(-1)^{n_X}C_i^{\rm ours},
\end{equation}
where $n_X$ is the number of field strengths in the operator. Operators built only from Higgs fields, fermions and covariant derivatives carry no relative sign.

For tensor fermion bilinears, we use $\sigma^{\mu\nu}=\tfrac{i}{2}[\gamma^\mu,\gamma^\nu]$, implemented as an explicit antisymmetrised product of two gamma matrices.

\subsection{The CP assignment}
\label{app:cp}

For the CP transformation defined in \cref{sec:cp}, we obtain the sign $\eta$ from the building blocks of the operator. A scalar fermion bilinear contributes $+1$, while a vector or tensor bilinear contributes $-1$, with or without derivatives. Each field strength contributes $-1$ and each dual field strength $+1$. The additional sign for a dual comes from the Levi-Civita tensor.

Gauge-index contractions also enter the assignment. Each $\epsilon^{IJK}$ or $f^{ABC}$ contributes $-1$, while $d^{ABC}$ and Kronecker deltas contribute $+1$. These signs follow from the transposition properties of the generators. For $SU(2)$, $(\tau^I)^T=s_I\tau^I$ with $s_1s_2s_3=-1$, and the corresponding relations hold for the Gell-Mann matrices. Higgs factors contribute $+1$ in the relation to the Hermitian-conjugate operator. An explicit factor of $i$ contributes $-1$.

We apply this rule to the corrected operator forms used in the models, including any factor of $i$ restored for Hermiticity. The resulting assignments are recorded in the coefficient dictionary and tested numerically as described in \cref{sec:validation}.

\section{Practical use and implementation checks}
\label{app:detail}

We begin with a short calculation using a released model. The remaining subsections describe the fixed-order extraction, comparisons with other implementations, removal of vanishing structures and operator dictionary.

\subsection{A short worked example}
\label{sec:example}

We calculate the interference of the $\psi^2X^2D$ class with the SM in $pp\to W^+W^-$. This example uses a released model, Python 3.11 and a working \MG\ installation. No \FR\ or Mathematica step is required. Run the shell commands from the root of the downloaded repository. First set the path to \MG\ and unpack the model:
\begin{lstlisting}
export MG5_DIR=/path/to/MG5_aMC
tar -xJf models/dim8_mw.tar.xz -C models
python3.11 "$MG5_DIR/bin/mg5_aMC"
\end{lstlisting}
At the \MG\ prompt, import the supplied class-14 restriction and generate the interference:
\begin{lstlisting}
import model ./models/dim8_mw-cls14
generate p p > w+ w- NP^2==2
output ww_cls14
quit
\end{lstlisting}
Back in the shell, set the parameters and generate events:
\begin{lstlisting}
python3 examples/studies/set_cards.py ww_cls14 \
    dim8_all=1 Lam=3000 MT=172 ymt=172 \
    ebeam1=6800 ebeam2=6800 nevents=100000 \
    dynamical_scale_choice=3 hel_recycling=False
python3.11 ww_cls14/bin/generate_events -f
\end{lstlisting}
These commands set the retained coefficients to one, $\Lambda=3$~TeV, the collider energy to $13.6$~TeV and both scales to $H_T/2$. Helicity recycling is disabled for compatibility with the dimension-eight Lorentz structures. The card-setting script rejects unknown parameter names. The stored benchmark gives an interference of approximately $-0.029$~pb, with independent runs fluctuating within their Monte Carlo uncertainties. The events are written to \code{ww\_cls14/Events/run\_01/}.

\begin{itemize}
\item Changing the process or operator selection.
To select all operators that can contribute to a process, use
\begin{lstlisting}
./dim8 select "p p > w+ w-" --card ww
\end{lstlisting}
For this example, the selector finds 95 operators through interaction vertices and six through the input relations alone. The restriction card retains 313 real dimension-eight coefficient entries. Import \code{./models/dim8\_mw-ww} to use it in place of the class-14 restriction. For another final state, change the process string in both the selection and generation commands and choose new card and output names. The optional command
\begin{lstlisting}
./dim8 count "psi^2X^2D"
\end{lstlisting}
reports the class's leading powers, $\lambda^4$ at four legs, $\lambda^6$ at five and $\lambda^8$ at six. Contributions that require light-fermion Yukawa couplings need a restriction card that retains those couplings, as explained in \cref{sec:select}.

\item Adding the SM and squared contribution.
Repeat the generation with \code{NP=0} and the output name \code{ww\_sm} for the SM reference. Use \code{NP<=2 NP\^{}2==4} and \code{ww\_cls14\_sq} for the squared single-insertion term. Apply the same parameters and run settings to each output directory before generating events. The stored benchmark gives approximately $68.4$~pb for the SM and $0.022$~pb for the square, corresponding to $\delta=-4.3\times10^{-4}$ and $R=0.74$ in \cref{tab:master}.

\item Automated studies and distributions.
The study driver can run the SM, interference and square together and retain the records used by the plotting scripts. A small study of the same class is configured by
\begin{lstlisting}
mkdir -p ww_study
printf "p p > w+ w-\n" > ww_study/processes.txt
printf "cls14\n" > ww_study/classes.txt
DIM8_ORDERS=NP=0 examples/studies/run_study.sh \
    ww_study models/dim8_mw 100000
\end{lstlisting}
The process list can contain several final states and their cuts, and the class list can contain several restriction cards. The driver saves the rates and recorded samples needed by the plotting tools, following the procedure in \cref{app:studies}. The plotting tools retain signed interference weights and report statistical uncertainties for mass bins, upper mass cuts and the ratio $R$. Bin uncertainties combine fluctuations in the weighted bin fraction with the integration error, retaining the correlation between the bin and the total event weight. The SM, interference and square are generated in separate runs and their errors are propagated independently. Plotting requires Matplotlib. The distribution study supplied with the repository gives the commands used for the figures in this paper.

\end{itemize}

\subsection{Fixed-order extraction in the $\{\alpha,M_Z,G_F\}$ scheme}
\label{app:extraction}

The extraction follows \cref{sec:inputs}. We hold the Wilson coefficients fixed, evaluate $X(t)$ at several values of $\Lambda$, and fit the coefficients of its expansion in $t=1/\Lambda^2$. The default scales, $\Lambda=2$, 3 and 4~TeV, are suitable for the produced-$W$ and off-shell benchmarks studied here. Changing the fitted scales tests the sensitivity to higher-order terms. Near a resolved $W$ resonance, larger scales are needed to keep the shift in $M_W^2$ small compared with $M_W\Gamma_W$.

Within each scan, the runs use a common random seed. Repeating the scan with independent seeds gives an estimate of the statistical uncertainty from their spread. A scan is included in the fit only when all required runs are present. We compare the extracted coefficients with an independent fixed-order expansion for $u\bar d\to e^+\nu_e$ and $pp\to W^+W^-$, as described in \cref{sec:validation}.

\subsection{Comparisons with other implementations}

For the comparison with \textsc{SmeftFR}, we load its dimension-eight operators into our \FR\ model. Both implementations then use the same fields, covariant derivative and unitary gauge. This tests the operator definitions in a common convention. We evaluate vertices of the difference at random momenta satisfying momentum conservation, with Levi-Civita tensors expanded explicitly. Expressions containing dual field strengths can cancel through momentum conservation and the Schouten identity. The operators in $XH^4D^2$ also require the relative factor of $i$ discussed in \cref{sec:basis}, since \textsc{SmeftFR} implements the forms printed in the Murphy basis.

For the dimension-six comparison, we use the $U(3)^5$ \UFO\ models of \SMEFTsim\ version 3.0.2 and its default inputs. We set the CKM matrix to the identity so that the different CKM parameterizations coincide. For each coefficient, symmetric finite differences extract the derivative of every coupling at zero coefficient, cancelling quadratic and cubic terms. The two implementations can decompose a vertex differently into Lorentz structures, colour tensors and Dirac projectors. We therefore compare fully contracted vertices at common random momenta, colour vectors, polarization vectors and spinors.

The released validation tools also run the Lorentz and CP tests described in \cref{sec:validation}. Separate audits check the EFT order assigned to each coefficient-dependent coupling and the mass dimension of every dimension-eight vertex. These checks complement the comparisons with independent implementations.

\subsection{Vanishing structures}
\label{app:zeros}

The exported interactions can contain Lorentz structures that cancel identically but remain unsimplified in the \FR\ output. Such structures can cause \code{ALOHA} to generate routines that fail to compile. We identify candidates for removal by evaluating them at generic momenta satisfying momentum conservation, with generic wavefunctions. This distinguishes cancellations of the structure itself from zeros that require particular on-shell external states. For four-fermion vertices, the contractions include the relative signs assigned by \MG\ to crossed fermion pairings.

Cancellations can also involve several structures or vertices. We group structures with common external particles by their couplings and test their signed sums. The tests use generic parameter values, including nonzero light-fermion Yukawa couplings, and compare the summed vertices before and after removal at common numerical points. These comparisons provide numerical checks that the retained interactions reproduce the original export.

Some zeros have a direct physical interpretation. The vacuum part of $(H^\dag H)^2X\widetilde X$ is a theta term, so its three-boson interaction without an external Higgs boson vanishes as a total derivative. The four-gluon vacuum interaction of $\Q{G^2H^4}{1}$ instead cancels the correction generated by gluon-field normalization, as required by the input treatment in \cref{sec:inputs}.

The leading flavour structures of thirteen operators vanish. Twelve belong to the four-fermion field-strength families $l^4B$, $q^4B$, $e^4B$, $u^4B$ and $d^4B$. With identical flavour contractions, the product of the two currents is symmetric in the Lorentz indices contracted with the antisymmetric field strength. The thirteenth is $\Q{q^4H^2}{5}$, whose $\epsilon^{IJK}$ contracts two identical triplet currents. Additional MFV structures distinguish the currents and give nonzero vertices for all thirteen operators except $\Q{e^4B}{2}$, which has no vertex within the implemented flavour expansion.

\subsection{Operator dictionary}

The accompanying machine-readable operator dictionary lists each coefficient name, its contribution to the Lagrangian and its CP assignment. It is generated from the same operator definitions as the model parameter cards and is also supplied in \LaTeX\ form. Each basis contains 734 entries, including the 674 baryon-number-conserving operators used in the models. The combined source contains 749 because it includes both sets of fifteen operators in $\psi^2H^2D^3$. The dictionary also identifies the thirteen operators whose leading flavour structure vanishes and records the reason.

\section{Energy counting for representative processes}
\label{app:lambda}

We apply the counting rules of \cref{sec:lambda-amps} to four representative processes. The expansions show how different interactions can produce energy growth relative to the SM. They use the default geometric basis and the $\{M_W,M_Z,G_F\}$ input scheme, with both dimension-six and dimension-eight operators.

We write $c_X$ for a generic Wilson coefficient in class $X$. The factors $\hat\Lambda^{-2}$ and $\hat\Lambda^{-4}$ label the operator dimension, with the physical suppression by $\Lambda$ already included in the powers of $\lambda$. The displayed expressions contain one EFT insertion. Numerical factors, Lorentz structures, universal input-parameter shifts and terms with two dimension-six insertions are not written separately. We set the light-fermion Yukawa couplings to zero and omit external fermion chiralities with no SM counterpart.

These are schematic expansions of allowed contributions. Gauge cancellations and helicity selection can remove a leading term from an amplitude or its interference. Terms with relative counting $E^4/\Lambda^4$ are highlighted in bold. Unless stated otherwise, external vector bosons are transverse.

\subsection{$q\bar q\to W^+W^-$}
\label{app:amp-ww}

After unpacking the combined model, the command \code{./dim8 series "u u~ > w+ w-"} gives the expansion. For transverse $W$ bosons, the SM contributes through $s$-channel and $t$-channel exchange, while the SMEFT also supplies a contact interaction:
\begin{align}
\mathcal{A}_s &= \lambda^{0}\, g_{\bar qqV}\, g_{VVV}
 + \lambda^{2}\, g_{\bar qqV}\, c_{X^3}/\hat\Lambda^2
 + \lambda^{4}\, \big(g_{\bar qqV}\, c_{X^2H^2} + g_{VVV}\, c_{\psi^2H^2D}\big)/\hat\Lambda^2 \nonumber\\
 &\quad + \lambda^{6}\, g_{\bar qqV}\, c_{X^3H^2}/\hat\Lambda^4
 + \lambda^{8}\, \big[g_{\bar qqV}\, (c_{X^2H^4} + c_{XH^4D^2}) + g_{VVV}\, c_{\psi^2H^4D}\big]/\hat\Lambda^4, \nonumber\\
\mathcal{A}_t &= \lambda^{0}\, g_{\bar qqV}^2
 + \lambda^{4}\, g_{\bar qqV}\, c_{\psi^2H^2D}/\hat\Lambda^2
 + \lambda^{8}\, g_{\bar qqV}\, c_{\psi^2H^4D}/\hat\Lambda^4, \nonumber\\
\mathcal{A}_c &= \lambda^{4}\, \boldsymbol{c_{\psi^2X^2D}}/\hat\Lambda^4
 + \lambda^{6}\, \big(c_{\psi^2XH^2D} + c_{\psi^2H^2D^3}\big)/\hat\Lambda^4.
\label{eq:amp-ww}
\end{align}
The unsuppressed SM amplitude is of order $\lambda^0$. At dimension six, the $X^3$ term at $\lambda^2$ therefore grows as $E^2/\Lambda^2$, while the terms at $\lambda^4$ scale as $v^2/\Lambda^2$. At dimension eight, the strongest growth comes from the $\psi^2X^2D$ contact interaction at $\lambda^4$, highlighted in bold. It scales as $E^4/\Lambda^4$ and requires no additional SM vertex. The contact terms at $\lambda^6$ scale as $E^2v^2/\Lambda^4$, while the terms at $\lambda^8$ have no energy enhancement.

These powers describe the amplitudes. Their interference with the SM also depends on helicity. In particular, the leading transverse amplitudes generated by $X^3$ do not interfere with the SM in the massless limit, so their amplitude growth does not directly translate into the same enhancement of the interference~\cite{Azatov:2016sqh}.

For $W_LW_L$, the full SM amplitude remains of order $\lambda^0$ after the gauge cancellations. The Higgs-current class $\psi^2H^2D$ now contributes at $\lambda^2$, giving $E^2/\Lambda^2$ growth. At dimension eight, $\psi^2H^2D^3$ contributes at $\lambda^4$, giving $E^4/\Lambda^4$ growth. The classes $\psi^2H^4D$ and $XH^4D^2$ contribute at $\lambda^6$, corresponding to $E^2v^2/\Lambda^4$. Thus the leading dimension-eight growth comes from different classes in the transverse and longitudinal channels. Resolving the polarization can help distinguish the interactions responsible for a high-energy deviation.

The assignment to individual classes follows the chosen operator basis. In the default basis, $\psi^2XH^2D$ and $\psi^2H^2D^3$ have no three-point vertices and contribute to the transverse amplitude only through their contact terms. The Murphy basis gives a different distribution between contact and exchange contributions.\footnote{In the Murphy basis, the $\psi^2H^2D^3$ class includes a $q\bar qZ$ vertex proportional to the squared momentum $k^2$ carried by the $Z$. In neutral-current Drell--Yan production, the decomposition $k^2/(k^2-M_Z^2)=1+M_Z^2/(k^2-M_Z^2)$ produces a local contribution at $\lambda^6$ and an exchange contribution at $\lambda^8$.}

\subsection{$q\bar q\to e^+e^-$}

The SM contributes through neutral-boson exchange, while four-fermion operators also give a contact interaction:
\begin{align}
\mathcal{A}_s &= \lambda^{0}\, g_{\bar qqV}\, g_{\bar eeV}
 + \lambda^{4}\, \big(g_{\bar qqV} + g_{\bar eeV}\big)\, c_{\psi^2H^2D}/\hat\Lambda^2
 + \lambda^{8}\, \big(g_{\bar qqV} + g_{\bar eeV}\big)\, c_{\psi^2H^4D}/\hat\Lambda^4, \nonumber\\
\mathcal{A}_c &= \lambda^{2}\, c_{\psi^4}/\hat\Lambda^2
 + \lambda^{4}\, \boldsymbol{c_{\psi^4D^2}}/\hat\Lambda^4
 + \lambda^{6}\, c_{\psi^4H^2}/\hat\Lambda^4.
\label{eq:amp-ee}
\end{align}
The leading contact contributions scale as $E^2/\Lambda^2$ for $\psi^4$ and $E^4/\Lambda^4$ for $\psi^4D^2$. The two derivatives in the latter supply the additional powers of energy. In $\psi^4H^2$, replacing the Higgs fields by their vacuum values instead gives $E^2v^2/\Lambda^4$. The displayed current-operator terms shift the neutral-current couplings without energy enhancement. The class $\psi^2H^2D^3$ is absent because it has no three-point vertex in the default basis.

\subsection{$qq'\to qq'HH$}

Higgs-pair production with two jets illustrates how additional external particles expose interactions with fewer vacuum insertions. For example, $\psi^4H^2$ gives $E^2v^2/\Lambda^4$ in a four-fermion amplitude, but its six-point contact interaction contributes at relative order $E^4/\Lambda^4$ in this process.

We separate contributions with a bosonic $VV\to HH$ core from those involving EFT interactions on the quark lines or a six-point contact. Two SM quark currents and two propagators dress the core with $\lambda^{-4}\lambda^8=\lambda^4$. We first consider transverse vectors in the core. To keep the expansion compact, let $\mathcal C_X$ denote the relevant linear combinations of coefficients in class $X$, including their SM coupling factors. With $a_{\rm SM}^{\rm core}$ denoting the coefficient of the leading SM power, representative terms are
\begin{align}
\mathcal A_{\rm core} &\sim \lambda^4\Bigg[
a_{\rm SM}^{\rm core}
+\lambda^2\frac{\mathcal C_{X^2H^2}}{\hat\Lambda^2}
+\lambda^4\left(\frac{\mathcal C_{H^4D^2}}{\hat\Lambda^2}
+\frac{\boldsymbol{\mathcal C_{X^2H^2D^2}}}{\hat\Lambda^4}\right) \nonumber\\
&\qquad +\frac{\lambda^6}{\hat\Lambda^4}
\left(\mathcal C_{H^4D^4}+\mathcal C_{X^2H^4}+\mathcal C_{XH^4D^2}\right)
+\lambda^8\frac{\mathcal C_{H^6D^2}}{\hat\Lambda^4}+\cdots\Bigg], \nonumber\\
\mathcal A_{\rm line} &\sim \lambda^6\frac{\mathcal C_{\psi^2H^2D}}{\hat\Lambda^2}
+\frac{\lambda^8}{\hat\Lambda^4}
\left(\boldsymbol{\mathcal C_{\psi^2H^2D^3}}
+\boldsymbol{\mathcal C_{\psi^2XH^2D}}
+\boldsymbol{\mathcal C_{\psi^4H^2}}\right)
+\lambda^{10}\frac{\mathcal C_{\psi^2H^4D}}{\hat\Lambda^4}+\cdots.
\label{eq:amp-hhjj}
\end{align}
The SM core includes both the $VVHH$ contact interaction and exchange diagrams. In Feynman gauge, the latter include Goldstone exchange at the leading power. The displayed core therefore represents their combined counting, rather than a contact interaction alone. Subleading terms include Higgs exchange and contributions from the potential classes $H^6$ and $H^8$.

The core class $X^2H^2$ and the quark-current class $\psi^2H^2D$ both give relative $E^2/\Lambda^2$ growth. At dimension eight, the four classes highlighted in bold give relative $E^4/\Lambda^4$ growth. Three involve the external quark lines, including the six-point interaction of $\psi^4H^2$.

The longitudinal channel supplies additional energy-enhanced contributions. Here a polarization assignment refers to the on-shell $VV\to HH$ core in the effective-vector-boson approximation. Applying the equivalence theorem to this core places $H^4D^2$ at $\lambda^2$, $H^4D^4$ at $\lambda^4$ and $H^6D^2$ at $\lambda^6$, before the quark-current dressing. Thus $H^4D^4$ can produce relative $E^4/\Lambda^4$ growth through longitudinal bosons, while $X^2H^2D^2$ supplies this growth in the transverse channel. The counting identifies these mechanisms without assuming that the unpolarized sample is dominated by longitudinal bosons.

\subsection{$u\bar d\to W^+W^-W^+$}

The SM amplitude begins at $\lambda^2$. Its leading topologies include successive gauge-boson emissions from the quark line, diagrams with triple-gauge vertices, and a quartic-gauge vertex connected to the quark current. Using $\mathcal C_X$ as above to include the topology-dependent SM coupling factors, the leading class powers are
\begin{align}
\mathcal A \sim{}& \lambda^2 a_{\rm SM}
+\lambda^4\frac{\mathcal C_{X^3}}{\hat\Lambda^2}
+\lambda^6\left[
\frac{\mathcal C_{X^2H^2}+\mathcal C_{\psi^2H^2D}}{\hat\Lambda^2}
+\frac{\boldsymbol{\mathcal C_{X^4}}+\boldsymbol{\mathcal C_{\psi^2X^2D}}}{\hat\Lambda^4}
\right] \nonumber\\
&+\frac{\lambda^8}{\hat\Lambda^4}
\left(\mathcal C_{X^3H^2}+\mathcal C_{X^2H^2D^2}
+\mathcal C_{\psi^2H^2D^3}+\mathcal C_{\psi^2XH^2D}\right) \nonumber\\
&+\frac{\lambda^{10}}{\hat\Lambda^4}
\left(\mathcal C_{H^4D^4}+\mathcal C_{X^2H^4}
+\mathcal C_{XH^4D^2}+\mathcal C_{\psi^2H^4D}\right)+\cdots,
\label{eq:amp-www}
\end{align}
where $a_{\rm SM}$ is the coefficient of the leading SM power. The quark-current factors in $\mathcal C_X$ include the charged-current flavour structure appropriate to the incoming $u\bar d$ state.

The dimension-six class $X^3$ gives relative $E^2/\Lambda^2$ growth. The classes $X^2H^2$ and $\psi^2H^2D$ enter at $\lambda^6$ and have no corresponding energy enhancement. At dimension eight, $X^4$ and $\psi^2X^2D$ lead at relative order $E^4/\Lambda^4$. The former contributes through a four-boson interaction, while the latter contributes through $q\bar qWW$ interactions and five-point contact terms. The additional final-state boson allows $X^4$ to interfere with the SM at tree level in quark annihilation. In diboson production its gluonic contact contribution instead has no tree-level SM counterpart. Longitudinal polarizations require the separate counting described in \cref{app:amp-ww}.

\section{Complete numerical results}
\label{app:studies}

\Cref{tab:grid} gives the dimension-eight interference for all 21 operator classes in the twelve processes of \cref{sec:pheno}. The entries are the signed quantity $\delta$ of \cref{eq:delta-r}, in units of $10^{-4}$. Four columns appear in \cref{tab:master}. All results use $13.6$~TeV proton collisions, $\Lambda=3$~TeV and unit coefficients within one class at a time, with the flavour and input choices of \cref{sec:pheno}.

The Drell--Yan samples require $m_{ee}>200$~GeV, $p_T^\ell>10$~GeV, $|\eta_\ell|<2.5$ and $\Delta R(\ell,\ell)>0.4$. The photon channel additionally requires $p_T^\gamma>20$~GeV, $|\eta_\gamma|<2.5$ and $\Delta R(\gamma,\ell)>0.4$. The electroweak $HHjj$ sample is generated with \code{QCD=0}, $p_T^j>25$~GeV, $|\eta_j|<5$, $m_{jj}>400$~GeV and $\Delta R(j,j)>0.4$. No kinematic cuts are applied to the diboson or triboson samples.

\begin{table}[H]
\centering
\scriptsize
\setlength{\tabcolsep}{2.6pt}
\resizebox{\textwidth}{!}{%
\begin{tabular}{@{}r l rrrrrrrrrrrr@{}}
\toprule
 & class & $W^+W^-$ & $W^+Z$ & $W^-Z$ & $ZZ$ & $W^+W^-W^+$ & $W^+W^-W^-$ & $W^+W^-Z$ & $W^+ZZ$ & $ZZZ$ & $HHjj$ & $e^+e^-$ & $e^+e^-\gamma$ \\
\cmidrule(lr){3-14}
 & $\sigma_{\rm SM}$ [pb] & $68.4$ & $16.3$ & $9.88$ & $9.89$ & $0.0786$ & $0.0425$ & $0.0897$ & $0.0189$ & $0.0098$ & $8.7\!\times\!10^{-4}$ & $1.25$ & $0.0193$ \\
\midrule
\rowcolor{black!8}\textbf{1} & $\boldsymbol{X^4}$ & $\ast$ & -- & -- & $\ast$ & $+16.4$ & $+12.9$ & $+3.13$ & $+11.4$ & $+33.8$ & -- & -- & $\ast$ \\
2 & $H^8$ & -- & -- & -- & -- & -- & -- & -- & -- & -- & $+5.77$ & -- & -- \\
3 & $H^6D^2$ & $-0.01$ & $+0.18$ & $+0.18$ & $+0.35$ & $-0.01$ & $-0.01$ & $+0.22$ & $+0.34$ & $+0.57$ & $-0.30$ & $-0.60$ & $-1.49$ \\
4 & $H^4D^4$ & -- & -- & -- & -- & $-0.03$ & $-0.03$ & $-0.01$ & $-0.03$ & $-0.11$ & $+11.4$ & -- & -- \\
5 & $X^3H^2$ & $-0.01$ & $-0.08$ & $-0.08$ & -- & $-0.02$ & $-0.03$ & $+0.58$ & $-0.15$ & -- & -- & -- & -- \\
6 & $X^2H^4$ & $-0.06$ & $+1.15$ & $+1.15$ & $+2.05$ & $+0.44$ & $+0.43$ & $+1.29$ & $+2.22$ & $+4.34$ & $-13.3$ & $-0.98$ & $-3.24$ \\
\rowcolor{black!8}\textbf{7} & $\boldsymbol{X^2H^2D^2}$ & $\ast$ & -- & -- & $\ast$ & $-0.44$ & $-0.36$ & $-0.09$ & $-0.87$ & $-1.76$ & $+8.52$ & -- & -- \\
8 & $XH^4D^2$ & $-0.01$ & $-0.06$ & $-0.06$ & -- & $-0.09$ & $-0.07$ & $-0.07$ & $-0.17$ & $-0.19$ & $-0.54$ & -- & -- \\
9 & $\psi^2X^2H$ & -- & -- & -- & -- & -- & -- & -- & -- & -- & -- & -- & -- \\
10 & $\psi^2XH^3$ & -- & -- & -- & -- & -- & -- & -- & -- & -- & -- & -- & -- \\
11 & $\psi^2H^2D^3$ & $-0.13$ & $-1.41$ & $-1.05$ & $+0.68$ & $-8.49$ & $-5.86$ & $-8.01$ & $-18.9$ & $-13.8$ & $-1.10$ & -- & -- \\
12 & $\psi^2H^5$ & -- & -- & -- & -- & -- & -- & -- & -- & -- & -- & -- & -- \\
13 & $\psi^2H^4D$ & $+0.19$ & $+1.19$ & $+1.16$ & $+0.64$ & $+2.12$ & $+1.99$ & $+1.49$ & $+3.92$ & $+6.18$ & $+16.5$ & $-0.51$ & $-1.58$ \\
\rowcolor{black!8}\textbf{14} & $\boldsymbol{\psi^2X^2D}$ & $-4.30$ & $-10.4$ & $-7.84$ & $-5.94$ & $-112$ & $-78.4$ & $-198$ & $-240$ & $-45.3$ & -- & $\ast$ & $-8.20$ \\
15 & $\psi^2XH^2D$ & $+0.11$ & $+0.83$ & $+0.73$ & $-0.27$ & $+2.65$ & $+2.25$ & $+0.58$ & $+5.84$ & $+7.18$ & $+11.9$ & -- & $+0.04$ \\
16 & $\psi^2XHD^2$ & -- & -- & -- & -- & -- & -- & -- & -- & -- & -- & -- & -- \\
17 & $\psi^2H^3D^2$ & -- & -- & -- & -- & -- & -- & -- & -- & -- & -- & -- & -- \\
18 & $\psi^4H^2$ & $+0.45$ & $+0.45$ & $+0.45$ & $+0.45$ & $+0.68$ & $+0.68$ & $+0.68$ & $+0.68$ & $+0.68$ & $+22.5$ & $+4.55$ & $+2.62$ \\
19 & $\psi^4X$ & -- & -- & -- & -- & -- & -- & -- & -- & -- & -- & -- & $-2.09$ \\
20 & $\psi^4HD$ & -- & -- & -- & -- & -- & -- & -- & -- & -- & -- & -- & -- \\
\rowcolor{black!8}\textbf{21} & $\boldsymbol{\psi^4D^2}$ & -- & -- & -- & -- & -- & -- & -- & -- & -- & -- & $+20.8$ & $-40.6$ \\
\bottomrule
\end{tabular}}
\caption{Dimension-eight interference in the twelve benchmark processes, in units of $10^{-4}$ of the corresponding SM rate. The first row gives the SM cross sections in pb. The Monte Carlo error of every entry is below five per cent. An asterisk denotes a nonzero squared contribution without tree-level SM interference, and a dash denotes no contributing diagram. The shaded classes are discussed in the main text. The settings and cuts are given in this appendix.}
\label{tab:grid}
\end{table}

The accompanying study results provide absolute cross sections, Monte Carlo integration errors, squared contributions and the $W$ mass used in each run. \MG\ estimates the integration error from the absolute values of the event weights, which understates the error of an interference, whose weights take both signs. For \cref{tab:master,tab:grid} we instead take the integration error and the error of the mean of the signed event weights added in quadrature, an upper bound on the error. The run records identify the model version, generator version, restriction card, parameter and run cards, and event sample. The table and figure scripts check these records and their consistency with the released model before using the results. The study driver retains the generated event files. Separate per-operator results for $pp\to W^+W^-$ cover the 95 operators with a vertex in the process and the six entering through the input relations alone.

\bibliographystyle{JHEP}
\bibliography{ref}

@article{Buchmuller:1985jz,
    author = "Buchmuller, W. and Wyler, D.",
    title = "{Effective Lagrangian Analysis of New Interactions and Flavor Conservation}",
    reportNumber = "CERN-TH-4254/85",
    doi = "10.1016/0550-3213(86)90262-2",
    journal = "Nucl. Phys. B",
    volume = "268",
    pages = "621--653",
    year = "1986"
}

@article{Grzadkowski:2010es,
    author = "Grzadkowski, B. and Iskrzynski, M. and Misiak, M. and Rosiek, J.",
    title = "{Dimension-Six Terms in the Standard Model Lagrangian}",
    eprint = "1008.4884",
    archivePrefix = "arXiv",
    primaryClass = "hep-ph",
    reportNumber = "IFT-9-2010, TTP10-35",
    doi = "10.1007/JHEP10(2010)085",
    journal = "JHEP",
    volume = "10",
    pages = "085",
    year = "2010"
}

@article{Contino:2016jqw,
    author = "Contino, Roberto and Falkowski, Adam and Goertz, Florian and Grojean, Christophe and Riva, Francesco",
    title = "{On the Validity of the Effective Field Theory Approach to SM Precision Tests}",
    eprint = "1604.06444",
    archivePrefix = "arXiv",
    primaryClass = "hep-ph",
    reportNumber = "DESY-16-067, CERN-TH-2016-082, LPT-Orsay-16-32",
    doi = "10.1007/JHEP07(2016)144",
    journal = "JHEP",
    volume = "07",
    pages = "144",
    year = "2016"
}

@article{Brivio:2017vri,
    author = "Brivio, Ilaria and Trott, Michael",
    title = "{The Standard Model as an Effective Field Theory}",
    eprint = "1706.08945",
    archivePrefix = "arXiv",
    primaryClass = "hep-ph",
    doi = "10.1016/j.physrep.2018.11.002",
    journal = "Phys. Rept.",
    volume = "793",
    pages = "1--98",
    year = "2019"
}

@article{Chanowitz:1985hj,
    author = "Chanowitz, Michael S. and Gaillard, Mary K.",
    title = "{The TeV Physics of Strongly Interacting W's and Z's}",
    reportNumber = "LBL-19470, UCB-PTH-85/19",
    doi = "10.1016/0550-3213(85)90580-2",
    journal = "Nucl. Phys. B",
    volume = "261",
    pages = "379--431",
    year = "1985"
}

@article{Degrande:2011ua,
    author = "Degrande, Celine and Duhr, Claude and Fuks, Benjamin and Grellscheid, David and Mattelaer, Olivier and Reiter, Thomas",
    title = "{UFO - The Universal FeynRules Output}",
    eprint = "1108.2040",
    archivePrefix = "arXiv",
    primaryClass = "hep-ph",
    reportNumber = "CP3-11-25, IPHC-PHENO-11-04, IPPP-11-39, DCPT-11-78, MPP-2011-68",
    doi = "10.1016/j.cpc.2012.01.022",
    journal = "Comput. Phys. Commun.",
    volume = "183",
    pages = "1201--1214",
    year = "2012"
}

@article{Chala:2021pll,
    author = "Chala, Mikael and Guedes, Guilherme and Ramos, Maria and Santiago, Jose",
    title = "{Towards the renormalisation of the Standard Model effective field theory to dimension eight: Bosonic interactions I}",
    eprint = "2106.05291",
    archivePrefix = "arXiv",
    primaryClass = "hep-ph",
    doi = "10.21468/SciPostPhys.11.3.065",
    journal = "SciPost Phys.",
    volume = "11",
    pages = "065",
    year = "2021"
}

@article{DasBakshi:2022mwk,
    author = "Das Bakshi, Supratim and Chala, Mikael and D{\'\i}az-Carmona, {\'A}lvaro and Guedes, Guilherme",
    title = "{Towards the renormalisation of the Standard Model effective field theory to dimension eight: bosonic interactions II}",
    eprint = "2205.03301",
    archivePrefix = "arXiv",
    primaryClass = "hep-ph",
    doi = "10.1140/epjp/s13360-022-03194-5",
    journal = "Eur. Phys. J. Plus",
    volume = "137",
    number = "8",
    pages = "973",
    year = "2022"
}

@article{Boughezal:2024zqa,
    author = "Boughezal, Radja and Huang, Yingsheng and Petriello, Frank",
    title = "{Renormalization-group running of dimension-8 four-fermion operators in the SMEFT}",
    eprint = "2408.15378",
    archivePrefix = "arXiv",
    primaryClass = "hep-ph",
    doi = "10.1103/PhysRevD.110.116015",
    journal = "Phys. Rev. D",
    volume = "110",
    number = "11",
    pages = "116015",
    year = "2024"
}

@article{Alloul:2013bka,
    author = "Alloul, Adam and Christensen, Neil D. and Degrande, C{\'e}line and Duhr, Claude and Fuks, Benjamin",
    title = "{FeynRules  2.0 - A complete toolbox for tree-level phenomenology}",
    eprint = "1310.1921",
    archivePrefix = "arXiv",
    primaryClass = "hep-ph",
    reportNumber = "CERN-PH-TH-2013-239, MCNET-13-14, IPPP-13-71, DCPT-13-142, PITT-PACC-1308",
    doi = "10.1016/j.cpc.2014.04.012",
    journal = "Comput. Phys. Commun.",
    volume = "185",
    pages = "2250--2300",
    year = "2014"
}

@article{Brivio:2017btx,
    author = "Brivio, Ilaria and Jiang, Yun and Trott, Michael",
    title = "{The SMEFTsim package, theory and tools}",
    eprint = "1709.06492",
    archivePrefix = "arXiv",
    primaryClass = "hep-ph",
    doi = "10.1007/JHEP12(2017)070",
    journal = "JHEP",
    volume = "12",
    pages = "070",
    year = "2017"
}

@article{Azatov:2016sqh,
    author = "Azatov, Aleksandr and Contino, Roberto and Machado, Camila S. and Riva, Francesco",
    title = "{Helicity selection rules and noninterference for BSM amplitudes}",
    eprint = "1607.05236",
    archivePrefix = "arXiv",
    primaryClass = "hep-ph",
    reportNumber = "CERN-TH-2016-165",
    doi = "10.1103/PhysRevD.95.065014",
    journal = "Phys. Rev. D",
    volume = "95",
    number = "6",
    pages = "065014",
    year = "2017"
}

@article{Lehman:2015coa,
    author = "Lehman, Landon and Martin, Adam",
    title = "{Low-derivative operators of the Standard Model effective field theory via Hilbert series methods}",
    eprint = "1510.00372",
    archivePrefix = "arXiv",
    primaryClass = "hep-ph",
    doi = "10.1007/JHEP02(2016)081",
    journal = "JHEP",
    volume = "02",
    pages = "081",
    year = "2016"
}

@article{Henning:2015alf,
    author = "Henning, Brian and Lu, Xiaochuan and Melia, Tom and Murayama, Hitoshi",
    title = "{2, 84, 30, 993, 560, 15456, 11962, 261485, ...: Higher dimension operators in the SM EFT}",
    eprint = "1512.03433",
    archivePrefix = "arXiv",
    primaryClass = "hep-ph",
    reportNumber = "UCB-PTH-15-14, IPMU15-0207",
    doi = "10.1007/JHEP08(2017)016",
    journal = "JHEP",
    volume = "08",
    pages = "016",
    year = "2017",
    note = "[Erratum: JHEP 09, 019 (2019)]"
}

@article{Degrande:2023iob,
    author = "Degrande, C{\'e}line and Li, Hao-Lin",
    title = "{Impact of dimension-8 SMEFT operators on diboson productions}",
    eprint = "2303.10493",
    archivePrefix = "arXiv",
    primaryClass = "hep-ph",
    doi = "10.1007/JHEP06(2023)149",
    journal = "JHEP",
    volume = "06",
    pages = "149",
    year = "2023"
}

@article{Brivio:2021yjb,
    author = "Brivio, Ilaria and Dawson, Sally and de Blas, Jorge and Durieux, Gauthier and Savard, Pierre and Denner, Ansgar and Freitas, Ayres and Hays, Chris and Pecjak, Ben and Vicini, Alessandro",
    title = "{Electroweak input parameters}",
    eprint = "2111.12515",
    archivePrefix = "arXiv",
    primaryClass = "hep-ph",
    reportNumber = "CERN-LPCC-2021-002, CERN-LHCEFTWG-2021-001",
    month = "11",
    year = "2021"
}

@article{Corbett:2023yhk,
    author = "Corbett, Tyler and Martin, Adam",
    title = "{Higgs associated production with a vector decaying to two fermions in the geoSMEFT}",
    eprint = "2306.00053",
    archivePrefix = "arXiv",
    primaryClass = "hep-ph",
    doi = "10.21468/SciPostPhys.16.1.019",
    journal = "SciPost Phys.",
    volume = "16",
    number = "1",
    pages = "019",
    year = "2024"
}

@article{Ellis:2023zim,
    author = "Ellis, John and Mimasu, Ken and Zampedri, Francesca",
    title = "{Dimension-8 SMEFT analysis of minimal scalar field extensions of the Standard Model}",
    eprint = "2304.06663",
    archivePrefix = "arXiv",
    primaryClass = "hep-ph",
    reportNumber = "KCL-PH-TH/2023-18, CERN-TH-2023-038",
    doi = "10.1007/JHEP10(2023)051",
    journal = "JHEP",
    volume = "10",
    pages = "051",
    year = "2023"
}

@article{Helset:2022pde,
    author = "Helset, Andreas and Jenkins, Elizabeth E. and Manohar, Aneesh V.",
    title = "{Renormalization of the Standard Model Effective Field Theory from geometry}",
    eprint = "2212.03253",
    archivePrefix = "arXiv",
    primaryClass = "hep-ph",
    reportNumber = "CALT-TH-2022-041",
    doi = "10.1007/JHEP02(2023)063",
    journal = "JHEP",
    volume = "02",
    pages = "063",
    year = "2023"
}

@article{Assi:2023zid,
    author = "Assi, Beno{\^\i}t and Helset, Andreas and Manohar, Aneesh V. and Pag{\`e}s, Julie and Shen, Chia-Hsien",
    title = "{Fermion geometry and the renormalization of the Standard Model Effective Field Theory}",
    eprint = "2307.03187",
    archivePrefix = "arXiv",
    primaryClass = "hep-ph",
    reportNumber = "CALT-TH-2023-024, FERMILAB-PUB-23-362-T",
    doi = "10.1007/JHEP11(2023)201",
    journal = "JHEP",
    volume = "11",
    pages = "201",
    year = "2023"
}

@article{Martin:2023tvi,
    author = "Martin, Adam",
    title = "{A case study of SMEFT $ \mathcal{O}\left(1/{\Lambda}^4\right) $ effects in diboson processes: pp \textrightarrow{} W$^{\pm}$(\ensuremath{\ell}$^{\pm}$\ensuremath{\nu})\ensuremath{\gamma}}",
    eprint = "2312.09867",
    archivePrefix = "arXiv",
    primaryClass = "hep-ph",
    doi = "10.1007/JHEP05(2024)223",
    journal = "JHEP",
    volume = "05",
    pages = "223",
    year = "2024"
}

@article{Martin:2023fad,
    author = "Martin, Adam and Trott, Michael",
    title = "{More accurate $ \sigma \left(\mathcal{GG}\to h\right),\Gamma \left(h\to \mathcal{GG},\mathcal{AA},\overline{\Psi}\Psi \right) $ and Higgs width results via the geoSMEFT}",
    eprint = "2305.05879",
    archivePrefix = "arXiv",
    primaryClass = "hep-ph",
    doi = "10.1007/JHEP01(2024)170",
    journal = "JHEP",
    volume = "01",
    pages = "170",
    year = "2024"
}

@article{Remmen:2019cyz,
    author = "Remmen, Grant N. and Rodd, Nicholas L.",
    title = "{Consistency of the Standard Model Effective Field Theory}",
    eprint = "1908.09845",
    archivePrefix = "arXiv",
    primaryClass = "hep-ph",
    doi = "10.1007/JHEP12(2019)032",
    journal = "JHEP",
    volume = "12",
    pages = "032",
    year = "2019"
}

@article{Adams:2006sv,
    author = "Adams, Allan and Arkani-Hamed, Nima and Dubovsky, Sergei and Nicolis, Alberto and Rattazzi, Riccardo",
    title = "{Causality, analyticity and an IR obstruction to UV completion}",
    eprint = "hep-th/0602178",
    archivePrefix = "arXiv",
    reportNumber = "CERN-PH-TH-2006-033, HUTP-06-A0005",
    doi = "10.1088/1126-6708/2006/10/014",
    journal = "JHEP",
    volume = "10",
    pages = "014",
    year = "2006"
}

@article{Corbett:2023qtg,
    author = "Corbett, Tyler and Desai, Jay and {\'E}boli, O. J. P. and Gonzalez-Garcia, M. C. and Martines, Matheus and Reimitz, Peter",
    title = "{Impact of dimension-eight SMEFT operators in the electroweak precision observables and triple gauge couplings analysis in universal SMEFT}",
    eprint = "2304.03305",
    archivePrefix = "arXiv",
    primaryClass = "hep-ph",
    reportNumber = "YITP-SB-2023-04, FERMILAB-PUB-23-134-V, UWThPh 2023-13",
    doi = "10.1103/PhysRevD.107.115013",
    journal = "Phys. Rev. D",
    volume = "107",
    number = "11",
    pages = "115013",
    year = "2023"
}

@article{Boughezal:2022nof,
    author = "Boughezal, Radja and Huang, Yingsheng and Petriello, Frank",
    title = "{Exploring the SMEFT at dimension eight with Drell-Yan transverse momentum measurements}",
    eprint = "2207.01703",
    archivePrefix = "arXiv",
    primaryClass = "hep-ph",
    doi = "10.1103/PhysRevD.106.036020",
    journal = "Phys. Rev. D",
    volume = "106",
    number = "3",
    pages = "036020",
    year = "2022"
}

@article{Corbett:2019cwl,
    author = "Corbett, Tyler and Helset, Andreas and Trott, Michael",
    title = "{Ward Identities for the Standard Model Effective Field Theory}",
    eprint = "1909.08470",
    archivePrefix = "arXiv",
    primaryClass = "hep-ph",
    doi = "10.1103/PhysRevD.101.013005",
    journal = "Phys. Rev. D",
    volume = "101",
    number = "1",
    pages = "013005",
    year = "2020"
}

@article{Alwall:2014hca,
    author = "Alwall, J. and Frederix, R. and Frixione, S. and Hirschi, V. and Maltoni, F. and Mattelaer, O. and Shao, H. -S. and Stelzer, T. and Torrielli, P. and Zaro, M.",
    title = "{The automated computation of tree-level and next-to-leading order differential cross sections, and their matching to parton shower simulations}",
    eprint = "1405.0301",
    archivePrefix = "arXiv",
    primaryClass = "hep-ph",
    reportNumber = "CERN-PH-TH-2014-064, CP3-14-18, LPN14-066, MCNET-14-09, ZU-TH-14-14",
    doi = "10.1007/JHEP07(2014)079",
    journal = "JHEP",
    volume = "07",
    pages = "079",
    year = "2014"
}

@article{Talbert:2021iqn,
    author = "Talbert, Jim and Trott, Michael",
    title = "{Dirac masses and mixings in the (geo)SM(EFT) and beyond}",
    eprint = "2107.03951",
    archivePrefix = "arXiv",
    primaryClass = "hep-ph",
    doi = "10.1007/JHEP11(2021)009",
    journal = "JHEP",
    volume = "11",
    pages = "009",
    year = "2021"
}

@article{Hays:2020scx,
    author = "Hays, Chris and Helset, Andreas and Martin, Adam and Trott, Michael",
    title = "{Exact SMEFT formulation and expansion to $\mathcal{O}(v^4/\Lambda^4)$}",
    eprint = "2007.00565",
    archivePrefix = "arXiv",
    primaryClass = "hep-ph",
    doi = "10.1007/JHEP11(2020)087",
    journal = "JHEP",
    volume = "11",
    pages = "087",
    year = "2020"
}

@article{Dawson:2021xei,
    author = "Dawson, Sally and Homiller, Samuel and Sullivan, Matthew",
    title = "{Impact of dimension-eight SMEFT contributions: A case study}",
    eprint = "2110.06929",
    archivePrefix = "arXiv",
    primaryClass = "hep-ph",
    doi = "10.1103/PhysRevD.104.115013",
    journal = "Phys. Rev. D",
    volume = "104",
    number = "11",
    pages = "115013",
    year = "2021"
}

@article{Corbett:2021eux,
    author = "Corbett, Tyler and Helset, Andreas and Martin, Adam and Trott, Michael",
    title = "{EWPD in the SMEFT to dimension eight}",
    eprint = "2102.02819",
    archivePrefix = "arXiv",
    primaryClass = "hep-ph",
    doi = "10.1007/JHEP06(2021)076",
    journal = "JHEP",
    volume = "06",
    pages = "076",
    year = "2021"
}

@article{Alioli:2020kez,
    author = "Alioli, Simone and Boughezal, Radja and Mereghetti, Emanuele and Petriello, Frank",
    title = "{Novel angular dependence in Drell-Yan lepton production via dimension-8 operators}",
    eprint = "2003.11615",
    archivePrefix = "arXiv",
    primaryClass = "hep-ph",
    reportNumber = "LA-UR-20-22498",
    doi = "10.1016/j.physletb.2020.135703",
    journal = "Phys. Lett. B",
    volume = "809",
    pages = "135703",
    year = "2020"
}

@article{Martin:2021cvs,
    author = "Martin, Adam and Trott, Michael",
    title = "{$ggh$ variations}",
    eprint = "2109.05595",
    archivePrefix = "arXiv",
    primaryClass = "hep-ph",
    doi = "10.1103/PhysRevD.105.076004",
    journal = "Phys. Rev. D",
    volume = "105",
    number = "7",
    pages = "076004",
    year = "2022"
}

@article{Trott:2021vqa,
    author = "Trott, Michael",
    title = "{Methodology for theory uncertainties in the standard model effective field theory}",
    eprint = "2106.13794",
    archivePrefix = "arXiv",
    primaryClass = "hep-ph",
    doi = "10.1103/PhysRevD.104.095023",
    journal = "Phys. Rev. D",
    volume = "104",
    number = "9",
    pages = "095023",
    year = "2021"
}

@article{Brivio:2022pyi,
    author = "Brivio, Ilaria and others",
    title = "{Truncation, validity, uncertainties}",
    eprint = "2201.04974",
    archivePrefix = "arXiv",
    primaryClass = "hep-ph",
    reportNumber = "CERN-LHCEFTWG-2021-002, CERN-LPCC-2022-01",
    month = "1",
    year = "2022"
}

@article{Zhang:2021eeo,
    author = "Zhang, Cen",
    title = "{SMEFTs living on the edge: determining the UV theories from positivity and extremality}",
    eprint = "2112.11665",
    archivePrefix = "arXiv",
    primaryClass = "hep-ph",
    doi = "10.1007/JHEP12(2022)096",
    journal = "JHEP",
    volume = "12",
    pages = "096",
    year = "2022"
}

@article{Corbett:2021cil,
    author = "Corbett, Tyler and Martin, Adam and Trott, Michael",
    title = "{Consistent higher order $ \sigma \left(\mathcal{GG}\to h\right) $, $ \Gamma \left(h\to \mathcal{GG}\right) $ and $\Gamma(h \to\gamma\gamma)$ in geoSMEFT}",
    eprint = "2107.07470",
    archivePrefix = "arXiv",
    primaryClass = "hep-ph",
    doi = "10.1007/JHEP12(2021)147",
    journal = "JHEP",
    volume = "12",
    pages = "147",
    year = "2021"
}

@article{Boughezal:2021tih,
    author = "Boughezal, Radja and Mereghetti, Emanuele and Petriello, Frank",
    title = "{Dilepton production in the SMEFT at $\mathcal{O}(1/\Lambda^4)$}",
    eprint = "2106.05337",
    archivePrefix = "arXiv",
    primaryClass = "hep-ph",
    reportNumber = "LA-UR-21-25379",
    doi = "10.1103/PhysRevD.104.095022",
    journal = "Phys. Rev. D",
    volume = "104",
    number = "9",
    pages = "095022",
    year = "2021"
}

@article{Hays:2018zze,
    author = "Hays, Chris and Martin, Adam and Sanz, Ver{\'o}nica and Setford, Jack",
    title = "{On the impact of dimension-eight SMEFT operators on Higgs measurements}",
    eprint = "1808.00442",
    archivePrefix = "arXiv",
    primaryClass = "hep-ph",
    doi = "10.1007/JHEP02(2019)123",
    journal = "JHEP",
    volume = "02",
    pages = "123",
    year = "2019"
}

@article{Corbett:2020bqv,
    author = "Corbett, Tyler",
    title = "{The Feynman rules for the SMEFT in the background field gauge}",
    eprint = "2010.15852",
    archivePrefix = "arXiv",
    primaryClass = "hep-ph",
    doi = "10.1007/JHEP03(2021)001",
    journal = "JHEP",
    volume = "03",
    pages = "001",
    year = "2021"
}

@article{Brivio:2020onw,
    author = "Brivio, Ilaria",
    title = "{SMEFTsim 3.0 {\textemdash} a practical guide}",
    eprint = "2012.11343",
    archivePrefix = "arXiv",
    primaryClass = "hep-ph",
    doi = "10.1007/JHEP04(2021)073",
    journal = "JHEP",
    volume = "04",
    pages = "073",
    year = "2021"
}

@article{Helset:2020yio,
    author = "Helset, Andreas and Martin, Adam and Trott, Michael",
    title = "{The Geometric Standard Model Effective Field Theory}",
    eprint = "2001.01453",
    archivePrefix = "arXiv",
    primaryClass = "hep-ph",
    doi = "10.1007/JHEP03(2020)163",
    journal = "JHEP",
    volume = "03",
    pages = "163",
    year = "2020"
}

@article{Kim:2022amu,
    author = "Kim, Taegyun and Martin, Adam",
    title = "{Monolepton production in SMEFT to $ \mathcal{O} $(1/{\ensuremath{\Lambda}}$^{4}$) and beyond}",
    eprint = "2203.11976",
    archivePrefix = "arXiv",
    primaryClass = "hep-ph",
    doi = "10.1007/JHEP09(2022)124",
    journal = "JHEP",
    volume = "09",
    pages = "124",
    year = "2022"
}

@article{Li:2020gnx,
    author = "Li, Hao-Lin and Ren, Zhe and Shu, Jing and Xiao, Ming-Lei and Yu, Jiang-Hao and Zheng, Yu-Hui",
    title = "{Complete set of dimension-eight operators in the standard model effective field theory}",
    eprint = "2005.00008",
    archivePrefix = "arXiv",
    primaryClass = "hep-ph",
    doi = "10.1103/PhysRevD.104.015026",
    journal = "Phys. Rev. D",
    volume = "104",
    number = "1",
    pages = "015026",
    year = "2021"
}

@article{Murphy:2020rsh,
    author = "Murphy, Christopher W.",
    title = "{Dimension-8 operators in the Standard Model Effective Field Theory}",
    eprint = "2005.00059",
    archivePrefix = "arXiv",
    primaryClass = "hep-ph",
    doi = "10.1007/JHEP10(2020)174",
    journal = "JHEP",
    volume = "10",
    pages = "174",
    year = "2020"
}

@article{Assi:2024zap,
    author = "Assi, Beno{\^\i}t and Martin, Adam",
    title = "{Energy-enhanced dimension eight SMEFT effects in VBF Higgs production}",
    eprint = "2410.21563",
    archivePrefix = "arXiv",
    primaryClass = "hep-ph",
    reportNumber = "FERMILAB-PUB-24-0786-V",
    doi = "10.1007/JHEP02(2025)029",
    journal = "JHEP",
    volume = "02",
    pages = "029",
    year = "2025"
}

@article{Assi:2025zmp,
    author = "Assi, Beno{\^\i}t and Martin, Adam",
    title = "{Energy-enhanced expansion of the standard model effective field theory}",
    eprint = "2504.10617",
    archivePrefix = "arXiv",
    primaryClass = "hep-ph",
    reportNumber = "FERMILAB-PUB-25-0291-V",
    doi = "10.1103/s18f-xch8",
    journal = "Phys. Rev. D",
    volume = "112",
    number = "1",
    pages = "015024",
    year = "2025"
}

@article{Chala:2021cgt,
    author = "Chala, Mikael and D{\'\i}az-Carmona, {\'A}lvaro and Guedes, Guilherme",
    title = "{A Green{\textquoteright}s basis for the bosonic SMEFT to dimension 8}",
    eprint = "2112.12724",
    archivePrefix = "arXiv",
    primaryClass = "hep-ph",
    doi = "10.1007/JHEP05(2022)138",
    journal = "JHEP",
    volume = "05",
    pages = "138",
    year = "2022"
}

@article{Christensen:2008py,
    author = "Christensen, Neil D. and Duhr, Claude",
    title = "{FeynRules - Feynman rules made easy}",
    eprint = "0806.4194",
    archivePrefix = "arXiv",
    primaryClass = "hep-ph",
    reportNumber = "MSUHEP-080616, CP3-08-20",
    doi = "10.1016/j.cpc.2009.02.018",
    journal = "Comput. Phys. Commun.",
    volume = "180",
    pages = "1614--1641",
    year = "2009"
}

@article{Criado:2019ugp,
    author = "Criado, Juan Carlos",
    title = "{BasisGen: automatic generation of operator bases}",
    eprint = "1901.03501",
    archivePrefix = "arXiv",
    primaryClass = "hep-ph",
    doi = "10.1140/epjc/s10052-019-6769-5",
    journal = "Eur. Phys. J. C",
    volume = "79",
    number = "3",
    pages = "256",
    year = "2019"
}

@article{Darme:2023jdn,
    author = "Darm{\'e}, Luc and others",
    title = "{UFO 2.0: the {\textquoteleft}Universal Feynman Output{\textquoteright} format}",
    eprint = "2304.09883",
    archivePrefix = "arXiv",
    primaryClass = "hep-ph",
    reportNumber = "BONN-TH-2023-03, DESY-23-051, FERMILAB-PUB-23-138-T, KA-TP-06-2023,
  MCNET-23-06, P3H-23-023, TIF-UNIMI-2023-11",
    doi = "10.1140/epjc/s10052-023-11780-9",
    journal = "Eur. Phys. J. C",
    volume = "83",
    number = "7",
    pages = "631",
    year = "2023"
}

@article{DasBakshi:2026ief,
    author = "Das Bakshi, Supratim and Chala, Mikael and Ren, Zhe",
    title = "{Renormalization of the SMEFT to dimension eight: Fermionic interactions II}",
    eprint = "2606.19202",
    archivePrefix = "arXiv",
    primaryClass = "hep-ph",
    doi = "10.1007/JHEP10(2026)211",
    journal = "JHEP",
    volume = "10",
    pages = "211",
    year = "2026"
}

@article{Dedes:2019uzs,
    author = "Dedes, A. and Paraskevas, M. and Rosiek, J. and Suxho, K. and Trifyllis, L.",
    title = "{SmeftFR {\textendash} Feynman rules generator for the Standard Model Effective Field Theory}",
    eprint = "1904.03204",
    archivePrefix = "arXiv",
    primaryClass = "hep-ph",
    doi = "10.1016/j.cpc.2019.106931",
    journal = "Comput. Phys. Commun.",
    volume = "247",
    pages = "106931",
    year = "2020"
}

@article{Dedes:2023zws,
    author = "Dedes, A. and Rosiek, J. and Ryczkowski, M. and Suxho, K. and Trifyllis, L.",
    title = "{SmeftFR v3 {\textendash} Feynman rules generator for the Standard Model Effective Field Theory}",
    eprint = "2302.01353",
    archivePrefix = "arXiv",
    primaryClass = "hep-ph",
    doi = "10.1016/j.cpc.2023.108943",
    journal = "Comput. Phys. Commun.",
    volume = "294",
    pages = "108943",
    year = "2024"
}

@article{Degrande:2013kka,
    author = "Degrande, Celine",
    title = "{A basis of dimension-eight operators for anomalous neutral triple gauge boson interactions}",
    eprint = "1308.6323",
    archivePrefix = "arXiv",
    primaryClass = "hep-ph",
    doi = "10.1007/JHEP02(2014)101",
    journal = "JHEP",
    volume = "02",
    pages = "101",
    year = "2014"
}

@article{Degrande:2020evl,
    author = "Degrande, C{\'e}line and Durieux, Gauthier and Maltoni, Fabio and Mimasu, Ken and Vryonidou, Eleni and Zhang, Cen",
    title = "{Automated one-loop computations in the standard model effective field theory}",
    eprint = "2008.11743",
    archivePrefix = "arXiv",
    primaryClass = "hep-ph",
    reportNumber = "CERN-TH-2020-140, CP3-20-42",
    doi = "10.1103/PhysRevD.103.096024",
    journal = "Phys. Rev. D",
    volume = "103",
    number = "9",
    pages = "096024",
    year = "2021"
}

@article{Durieux:2024zrg,
    author = "Durieux, Gauthier and Remmen, Grant N. and Rodd, Nicholas L. and {\'E}boli, O. J. P. and Gonzalez-Garcia, Maria C. and Kondo, Dan and Murayama, Hitoshi and Okabe, Risshin",
    title = "{LHC EFT WG note: Basis for anomalous quartic gauge couplings}",
    eprint = "2411.02483",
    archivePrefix = "arXiv",
    primaryClass = "hep-ph",
    reportNumber = "CERN-LHCEFTWG-2024-002, CERN-LPCC-2024-002",
    doi = "10.21468/SciPostPhysCommRep.6",
    journal = "SciPost Phys. Comm. Rep.",
    pages = "6",
    year = "2025"
}

@article{Eboli:2006wa,
    author = "Eboli, O. J. P. and Gonzalez-Garcia, M. C. and Mizukoshi, J. K.",
    title = "{$pp \to jj e^\pm \mu^\pm \nu\nu$ and $jj e^\pm \mu^\mp \nu\nu$ at $\mathcal{O}(\alpha_{\rm em}^6)$ and $\mathcal{O}(\alpha_{\rm em}^4 \alpha_s^2)$ for the study of the quartic electroweak gauge boson vertex at CERN LHC}",
    eprint = "hep-ph/0606118",
    archivePrefix = "arXiv",
    reportNumber = "YITP-SB-06-10, IFUSP-1620-2006",
    doi = "10.1103/PhysRevD.74.073005",
    journal = "Phys. Rev. D",
    volume = "74",
    pages = "073005",
    year = "2006"
}

@article{Eboli:2016kko,
    author = "{\'E}boli, O. J. P. and Gonzalez-Garcia, M. C.",
    title = "{Classifying the bosonic quartic couplings}",
    eprint = "1604.03555",
    archivePrefix = "arXiv",
    primaryClass = "hep-ph",
    reportNumber = "YITP-SB-16-09",
    doi = "10.1103/PhysRevD.93.093013",
    journal = "Phys. Rev. D",
    volume = "93",
    number = "9",
    pages = "093013",
    year = "2016"
}

@article{ElFaham:2025fow,
    author = "El Faham, Hesham and Ventura, Giuseppe and Vryonidou, Eleni",
    title = "{Diboson production in the SMEFT at dimension-8}",
    eprint = "2511.04338",
    archivePrefix = "arXiv",
    primaryClass = "hep-ph",
    reportNumber = "TIF-UNIMI-2025-21",
    doi = "10.1007/JHEP04(2026)050",
    journal = "JHEP",
    volume = "04",
    pages = "050",
    year = "2026"
}

@article{Fonseca:2017lem,
    author = "Fonseca, Renato M.",
    editor = "Grzadkowski, Bohdan and Kalinowski, Jan and Krawczyk, Maria",
    title = "{The Sym2Int program: going from symmetries to interactions}",
    eprint = "1703.05221",
    archivePrefix = "arXiv",
    primaryClass = "hep-ph",
    doi = "10.1088/1742-6596/873/1/012045",
    journal = "J. Phys. Conf. Ser.",
    volume = "873",
    number = "1",
    pages = "012045",
    year = "2017"
}

@article{Fuentes-Martin:2022jrf,
    author = {Fuentes-Mart{\'\i}n, Javier and K{\"o}nig, Matthias and Pag{\`e}s, Julie and Thomsen, Anders Eller and Wilsch, Felix},
    title = "{A proof of concept for matchete: an automated tool for matching effective theories}",
    eprint = "2212.04510",
    archivePrefix = "arXiv",
    primaryClass = "hep-ph",
    reportNumber = "MITP-22-105, TUM-HEP-1443/22, ZU-TH-58/22",
    doi = "10.1140/epjc/s10052-023-11726-1",
    journal = "Eur. Phys. J. C",
    volume = "83",
    number = "7",
    pages = "662",
    year = "2023"
}

@article{Harlander:2023ozs,
    author = "Harlander, Robert V. and Schaaf, Magnus C.",
    title = "{AutoEFT: Automated operator construction for effective field theories}",
    eprint = "2309.15783",
    archivePrefix = "arXiv",
    primaryClass = "hep-ph",
    reportNumber = "TTK-23-25; P3H-23-066",
    doi = "10.1016/j.cpc.2024.109198",
    journal = "Comput. Phys. Commun.",
    volume = "300",
    pages = "109198",
    year = "2024"
}

@article{Marinissen:2020jmb,
    author = "Marinissen, Coenraad B. and Rahn, Rudi and Waalewijn, Wouter J.",
    title = "{..., 83106786, 114382724, 1509048322, 2343463290, 27410087742, ... efficient Hilbert series for effective theories}",
    eprint = "2004.09521",
    archivePrefix = "arXiv",
    primaryClass = "hep-ph",
    reportNumber = "NIKHEF 20-010",
    doi = "10.1016/j.physletb.2020.135632",
    journal = "Phys. Lett. B",
    volume = "808",
    pages = "135632",
    year = "2020"
}

@article{Hourtz:2026ijm,
    author = {Hourtz, Luis and Ketheeswaran, Tamilarasan and Kr{\"a}mer, Michael and Thomsen, Anders Eller and Weber, Philip and Wilsch, Felix},
    title = "{UFO Sightings with Matchete: Connecting BSM Lagrangians with Event Generators}",
    eprint = "2609.28619",
    archivePrefix = "arXiv",
    primaryClass = "hep-ph",
    reportNumber = "P3H-26-071, TTK-26-34",
    month = "9",
    year = "2026"
}

@article{Ellis:2018cos,
    author = "Ellis, John and Ge, Shao-Feng",
    title = "{Constraining Gluonic Quartic Gauge Coupling Operators with $gg \to \gamma\gamma$}",
    eprint = "1802.02416",
    archivePrefix = "arXiv",
    primaryClass = "hep-ph",
    reportNumber = "KCL-PH-TH-2018-02, CERN-TH-2018-014, IPMU-18-0022",
    doi = "10.1103/PhysRevLett.121.041801",
    journal = "Phys. Rev. Lett.",
    volume = "121",
    number = "4",
    pages = "041801",
    year = "2018"
}

@article{Brivio:2017bnu,
    author = "Brivio, Ilaria and Trott, Michael",
    title = "{Scheming in the SMEFT... and a reparameterization invariance!}",
    eprint = "1701.06424",
    archivePrefix = "arXiv",
    primaryClass = "hep-ph",
    doi = "10.1007/JHEP07(2017)148",
    journal = "JHEP",
    volume = "07",
    pages = "148",
    year = "2017",
    note = "[Addendum: JHEP 05, 136 (2018)]"
}

@article{Zhang:2018shp,
    author = "Zhang, Cen and Zhou, Shuang-Yong",
    title = "{Positivity bounds on vector boson scattering at the LHC}",
    eprint = "1808.00010",
    archivePrefix = "arXiv",
    primaryClass = "hep-ph",
    reportNumber = "USTC-ICTS-18-13",
    doi = "10.1103/PhysRevD.100.095003",
    journal = "Phys. Rev. D",
    volume = "100",
    number = "9",
    pages = "095003",
    year = "2019"
}

@article{Bi:2019phv,
    author = "Bi, Qi and Zhang, Cen and Zhou, Shuang-Yong",
    title = "{Positivity constraints on aQGC: carving out the physical parameter space}",
    eprint = "1902.08977",
    archivePrefix = "arXiv",
    primaryClass = "hep-ph",
    reportNumber = "USTC-ICTS-19-01",
    doi = "10.1007/JHEP06(2019)137",
    journal = "JHEP",
    volume = "06",
    pages = "137",
    year = "2019"
}

@article{Remmen:2020vts,
    author = "Remmen, Grant N. and Rodd, Nicholas L.",
    title = "{Flavor Constraints from Unitarity and Analyticity}",
    eprint = "2004.02885",
    archivePrefix = "arXiv",
    primaryClass = "hep-ph",
    doi = "10.1103/PhysRevLett.125.081601",
    journal = "Phys. Rev. Lett.",
    volume = "125",
    number = "8",
    pages = "081601",
    year = "2020",
    note = "[Erratum: Phys.Rev.Lett. 127, 149901 (2021)]"
}

@article{deAquino:2011ub,
    author = "de Aquino, Priscila and Link, William and Maltoni, Fabio and Mattelaer, Olivier and Stelzer, Tim",
    title = "{ALOHA: Automatic Libraries Of Helicity Amplitudes for Feynman Diagram Computations}",
    eprint = "1108.2041",
    archivePrefix = "arXiv",
    primaryClass = "hep-ph",
    reportNumber = "CP3-11-26",
    doi = "10.1016/j.cpc.2012.05.004",
    journal = "Comput. Phys. Commun.",
    volume = "183",
    pages = "2254--2263",
    year = "2012"
}

@article{Ball:2013hta,
    author = "Ball, Richard D. and Bertone, Valerio and Carrazza, Stefano and Del Debbio, Luigi and Forte, Stefano and Guffanti, Alberto and Hartland, Nathan P. and Rojo, Juan",
    collaboration = "NNPDF",
    title = "{Parton distributions with QED corrections}",
    eprint = "1308.0598",
    archivePrefix = "arXiv",
    primaryClass = "hep-ph",
    reportNumber = "EDINBURGH-2013-20, FR-PHENO-2013-008, CERN-PH-TH-2013-075, Edinburgh 2013/20, IFUM-1014-FT, FR-PHENO-2013-008,
  CERN-PH-TH/2013-075",
    doi = "10.1016/j.nuclphysb.2013.10.010",
    journal = "Nucl. Phys. B",
    volume = "877",
    pages = "290--320",
    year = "2013"
}

@article{Wulzer:2013mza,
    author = "Wulzer, Andrea",
    title = "{An Equivalent Gauge and the Equivalence Theorem}",
    eprint = "1309.6055",
    archivePrefix = "arXiv",
    primaryClass = "hep-ph",
    doi = "10.1016/j.nuclphysb.2014.05.021",
    journal = "Nucl. Phys. B",
    volume = "885",
    pages = "97--126",
    year = "2014"
}

@article{Cuomo:2019siu,
    author = "Cuomo, Gabriel and Vecchi, Luca and Wulzer, Andrea",
    title = "{Goldstone Equivalence and High Energy Electroweak Physics}",
    eprint = "1911.12366",
    archivePrefix = "arXiv",
    primaryClass = "hep-ph",
    reportNumber = "CERN-TH-2019-197",
    doi = "10.21468/SciPostPhys.8.5.078",
    journal = "SciPost Phys.",
    volume = "8",
    number = "5",
    pages = "078",
    year = "2020"
}

@article{Trott:2023jrw,
    author = "Trott, Michael",
    title = "{$\alpha_s$ as an input parameter in the SMEFT}",
    eprint = "2306.14784",
    archivePrefix = "arXiv",
    primaryClass = "hep-ph",
    month = "6",
    year = "2023"
}

@article{Assi:2026vjx,
    author = "Assi, Beno{\^\i}t and Martin, Adam and Shepherd, William",
    title = "{EFT validity and truncation uncertainty from few nuisance parameters}",
    eprint = "2607.02649",
    archivePrefix = "arXiv",
    primaryClass = "hep-ph",
    doi = "10.1007/JHEP11(2026)087",
    journal = "JHEP",
    volume = "11",
    pages = "087",
    year = "2026"
}

@article{Assi:2025fsm,
    author = "Assi, Beno{\^\i}t and Helset, Andreas and Pag{\`e}s, Julie and Shen, Chia-Hsien",
    title = "{Renormalizing two-fermion operators in the SMEFT via supergeometry}",
    eprint = "2504.18537",
    archivePrefix = "arXiv",
    primaryClass = "hep-ph",
    reportNumber = "CERN-TH-2025-084, FERMILAB-PUB-25-0280-V",
    doi = "10.1007/JHEP12(2025)082",
    journal = "JHEP",
    volume = "12",
    pages = "082",
    year = "2025"
}

@article{Alonso:2015fsp,
    author = "Alonso, Rodrigo and Jenkins, Elizabeth E. and Manohar, Aneesh V.",
    title = "{A Geometric Formulation of Higgs Effective Field Theory: Measuring the Curvature of Scalar Field Space}",
    eprint = "1511.00724",
    archivePrefix = "arXiv",
    primaryClass = "hep-ph",
    reportNumber = "CERN-PH-TH-2015-257",
    doi = "10.1016/j.physletb.2016.01.041",
    journal = "Phys. Lett. B",
    volume = "754",
    pages = "335--342",
    year = "2016"
}

@article{Alonso:2016oah,
    author = "Alonso, Rodrigo and Jenkins, Elizabeth E. and Manohar, Aneesh V.",
    title = "{Geometry of the Scalar Sector}",
    eprint = "1605.03602",
    archivePrefix = "arXiv",
    primaryClass = "hep-ph",
    reportNumber = "CERN-TH-2016-116",
    doi = "10.1007/JHEP08(2016)101",
    journal = "JHEP",
    volume = "08",
    pages = "101",
    year = "2016"
}

@article{Alonso:2021rac,
    author = "Alonso, Rodrigo and West, Mia",
    title = "{Roads to the Standard Model}",
    eprint = "2109.13290",
    archivePrefix = "arXiv",
    primaryClass = "hep-ph",
    doi = "10.1103/PhysRevD.105.096028",
    journal = "Phys. Rev. D",
    volume = "105",
    number = "9",
    pages = "096028",
    year = "2022"
}

@article{Cohen:2020xca,
    author = "Cohen, Timothy and Craig, Nathaniel and Lu, Xiaochuan and Sutherland, Dave",
    title = "{Is SMEFT Enough?}",
    eprint = "2008.08597",
    archivePrefix = "arXiv",
    primaryClass = "hep-ph",
    doi = "10.1007/JHEP03(2021)237",
    journal = "JHEP",
    volume = "03",
    pages = "237",
    year = "2021"
}

@article{Cohen:2021ucp,
    author = "Cohen, Timothy and Craig, Nathaniel and Lu, Xiaochuan and Sutherland, Dave",
    title = "{Unitarity violation and the geometry of Higgs EFTs}",
    eprint = "2108.03240",
    archivePrefix = "arXiv",
    primaryClass = "hep-ph",
    doi = "10.1007/JHEP12(2021)003",
    journal = "JHEP",
    volume = "12",
    pages = "003",
    year = "2021"
}

@article{Corbett:2020ymv,
    author = "Corbett, Tyler and Trott, Michael",
    title = "{One loop verification of SMEFT Ward Identities}",
    eprint = "2010.08451",
    archivePrefix = "arXiv",
    primaryClass = "hep-ph",
    doi = "10.21468/SciPostPhys.10.6.144",
    journal = "SciPost Phys.",
    volume = "10",
    number = "6",
    pages = "144",
    year = "2021"
}

@article{Gomez-Ambrosio:2022qsi,
    author = "G{\'o}mez-Ambrosio, Raquel and Llanes-Estrada, Felipe J. and Salas-Bern{\'a}rdez, Alexandre and Sanz-Cillero, Juan J.",
    title = "{Distinguishing electroweak EFTs with WLWL{\textrightarrow}n{\texttimes}h}",
    eprint = "2204.01763",
    archivePrefix = "arXiv",
    primaryClass = "hep-ph",
    doi = "10.1103/PhysRevD.106.053004",
    journal = "Phys. Rev. D",
    volume = "106",
    number = "5",
    pages = "053004",
    year = "2022"
}

@article{Gomez-Ambrosio:2022why,
    author = "G{\'o}mez-Ambrosio, Raquel and Llanes-Estrada, Felipe J. and Salas-Bern{\'a}rdez, Alexandre and Sanz-Cillero, Juan J.",
    title = "{SMEFT is falsifiable through multi-Higgs measurements (even in the absence of new light particles)}",
    eprint = "2207.09848",
    archivePrefix = "arXiv",
    primaryClass = "hep-ph",
    doi = "10.1088/1572-9494/ace95e",
    journal = "Commun. Theor. Phys.",
    volume = "75",
    number = "9",
    pages = "095202",
    year = "2023"
}

@article{Helset:2018fgq,
    author = "Helset, Andreas and Paraskevas, Michael and Trott, Michael",
    title = "{Gauge fixing the Standard Model Effective Field Theory}",
    eprint = "1803.08001",
    archivePrefix = "arXiv",
    primaryClass = "hep-ph",
    doi = "10.1103/PhysRevLett.120.251801",
    journal = "Phys. Rev. Lett.",
    volume = "120",
    number = "25",
    pages = "251801",
    year = "2018"
}

@article{Misiak:2018gvl,
    author = "Misiak, M. and Paraskevas, M. and Rosiek, J. and Suxho, K. and Zglinicki, B.",
    title = "{Effective Field Theories in R$_\xi$ gauges}",
    eprint = "1812.11513",
    archivePrefix = "arXiv",
    primaryClass = "hep-ph",
    doi = "10.1007/JHEP02(2019)051",
    journal = "JHEP",
    volume = "02",
    pages = "051",
    year = "2019"
}

@article{Salas-Bernardez:2022hqv,
    author = "Salas-Bernardez, Alexandre and Sanz-Cillero, Juan J. and Llanes-Estrada, Felipe J. and Gomez-Ambrosio, Raquel",
    title = "{SMEFT as a slice of HEFT{\textquoteright}s parameter space}",
    eprint = "2211.09605",
    archivePrefix = "arXiv",
    primaryClass = "hep-ph",
    doi = "10.1051/epjconf/202227408013",
    journal = "EPJ Web Conf.",
    volume = "274",
    pages = "08013",
    year = "2022"
}

@article{ParticleDataGroup:2024cfk,
    author = "Navas, S. and others",
    collaboration = "Particle Data Group",
    title = "{Review of particle physics}",
    doi = "10.1103/PhysRevD.110.030001",
    journal = "Phys. Rev. D",
    volume = "110",
    number = "3",
    pages = "030001",
    year = "2024"
}

@article{FCC:2025lpp,
    author = "Benedikt, M. and others",
    collaboration = "FCC",
    title = "{Future Circular Collider Feasibility Study Report: Volume 1, Physics, Experiments, Detectors}",
    eprint = "2505.00272",
    archivePrefix = "arXiv",
    primaryClass = "hep-ex",
    reportNumber = "CERN-FCC-PHYS-2025-0002",
    doi = "10.1140/epjc/s10052-025-15077-x",
    journal = "Eur. Phys. J. C",
    volume = "85",
    number = "12",
    pages = "1468",
    year = "2025",
    note = "[Erratum: Eur.Phys.J.C 86, 844 (2026)]"
}

@article{Gillies:2024mqp,
    author = "Gillies, Daniel and Banfi, Andrea and Martin, Adam and Lim, Matthew A.",
    title = "{Dimension-8 operators in W$^{+}$W$^{−}$ production via gluon fusion}",
    eprint = "2412.16020",
    archivePrefix = "arXiv",
    primaryClass = "hep-ph",
    doi = "10.1007/JHEP06(2025)111",
    journal = "JHEP",
    volume = "06",
    pages = "111",
    year = "2025"
}

@article{Gillies:2026tpc,
    author = "Gillies, Daniel and Banfi, Andrea and Martin, Adam",
    title = "{Probing EFT breakdown in the tails of $W^+ W^-$ observables}",
    eprint = "2601.19495",
    archivePrefix = "arXiv",
    primaryClass = "hep-ph",
    doi = "10.1140/epjc/s10052-026-16081-5",
    journal = "Eur. Phys. J. C",
    volume = "86",
    number = "9",
    pages = "1084",
    year = "2026"
}
\end{document}